\documentstyle[12pt,epsf,rotate]{article}

\newcommand \be  {\begin{equation}}
\newcommand \bea {\begin{eqnarray} \nonumber }
\newcommand \ee  {\end{equation}}
\newcommand \eea {\end{eqnarray}}

 \def\(({\left(}
 \def\)){\right)}
\def\[[{\left[}
\def\]]{\right]}

\def\ra{\rangle}
\def \a{\alpha}
\def\d{{\mbox d}}

\def\ov{\over}
\def\l{\left}
\def\r{\right}

\def\a{\alpha}
\def\b{\beta}

\def\de{\delta}

\def\vv{{\mbox{\boldmath{v}}}}
\def\cP{\cal P}
\def \bE{\bf E}
\def\fa{f_\alpha}
\def\fb{f_$$ta}
\def\ra{r_\alpha}
\def\rb{r_$$ta}
\def\l{\lambda}
\def\m{\mu}
\def \v{ {\vec v}}
\def \vx{\vec x}
\def \vk{\vec k}
\def \vf{\vec f}
\def \h{ {h(\vec x,t)}}
\def \p{ {\phi(\vec x,t)}}
\def \Z{ {Z(\vec x,t)}}
\def \V{ {V(\vec x,t)}}

\def \r{ {\rho(\vec x,t)}}
\def \dw{ {\int \tilde {d\omega}}}
\def \GG{ {G_{0}(\omega)}}
\def \Ga{ {G_{aa}(\omega)}}
\def \Gb{ {G_{bb}(\omega)}}
\def \Gab{ {G_{ab}(\omega)}}
\def \G1{ {G_{1}(\omega)}}

\def \Gc{ {G_{c}(\omega)}}
\def \Gt{ {\tilde G(\omega)}}

\def \f{ {\vec f(\vec k,t)}}
\def \e2{ {\eta(-\vec k,-\omega)}}
\def \dr{ {\delta \rho}}
\def \rp {{r_\perp}}
\def \d{ {\delta \rho(\vec x,t)}}

 \def\(({\left(}
 \def\)){\right)}
\def\[[{\left[}
\def\]]{\right]}

\def\ra{\rangle}
\def \a{\alpha}

\def\d{{\mbox d}}
\def\de{\delta}
\def\vv{{\mbox{\boldmath{v}}}}
\def\cP{\cal P}
\def \bE{\bf E}
\def\fa{f_\alpha}
\def\fb{f_$$ta}
\def\ra{r_\alpha}

\def\vra{\vec r_\alpha}
\def\rb{r_$$ta}
\def\l{\lambda}
\def\m{\mu}
\def\ml{|l|}

\begin{document}
\topmargin -1.0 true cm
\textheight 23 true cm
\baselineskip .5 true cm
\begin{titlepage}

\vskip .27in
\begin{center}

{\large \bf

SCALING AND INTERMITTENCY IN BURGERS' TURBULENCE

}

\vskip .27in

{J.P. Bouchaud }
\vskip .2in
{ \it Service
de Physique de l'Etat Condens\'e, CEA-Saclay, Orme des Merisiers, 91 191 Gif
s/ Yvette CEDEX, France}
\vskip .2in
{M. M\'ezard }
\vskip .2in
{\it Laboratoire de Physique Th\'eorique de l' Ecole Normale Sup\'erieure
 \footnote {Unit\'e propre du CNRS,  associ\'ee
 \`a\ l'Ecole
 Normale Sup\'erieure et \`a\ l'Universit\'e de Paris Sud} , 24 rue
 Lhomond, 75231 Paris Cedex 05, France}
\vskip .2in
{G. Parisi }
 \vskip .2in
{ \it Universita di Roma I, Piazza A.Moro 2, 00185 Roma, Italy}

\end{center}
\date\today
\input{psfig}

\vskip 8pt

{\bf Abstract}:
We use the mapping between Burgers' equation  and the problem of a directed
polymer in a random medium in order to study the fully developped
turbulence in the $N$ dimensional forced Burgers' equation. The
stirring force corresponds to a quenched (spatio temporal) random
potential for the polymer. The properties of the inertial regime
are deduced from a study of the directed polymer on length
scales smaller than the correlation length of the potential, which
is not the regime usually considered  in the case of polymers.
 From this study we
propose an Ansatz for the velocity field in the large
Reynolds number limit of the forced Burgers' equation in $N$ dimensions, which
should
become exact in the limit $N \to \infty$. This Ansatz allows us to compute
exactly the
full probability distribution of the velocity difference $u(r)$ between points
separated by a distance $r$ much smaller than the correlation length of the
forcing. We
find that the moments $<u^q(r)>$ scale as $r^{\zeta(q)}$ with $\zeta(q) \equiv
1$ for
all $q \geq 1$ (in particular the $q=3$ moment agrees with Kolmogorov's scaling
$\zeta(3)=1$). This strong `intermittency' is related to the large scale
singularities
of the velocity field, which is concentrated on a $N-1$ dimensional froth-like
structure, which
is turn related to the one-step replica symmetry broken nature of the
associated
disordered problem. We also discuss the similarities and differences between
Burgers
turbulence and hydrodynamical turbulence, and we comment on the
anomalous tracer fluctuations in a Burgers turbulent field.
This replica approach being rather unusual in turbulence problems,
we provide all the necessary details of the method.

\vfill
{ {\bf \footnotesize
LPTENS preprint 95/12}}
\vfill
\end{titlepage}

\newpage
\section{Introduction}

The theory of statistical turbulence is more than fifty years old, but its
status
is still not fully satisfactory. The simplest and surprisingly robust approach
is
Kolmogorov's dimensional analysis, which leads to the celebrated $k^{-5/3}$ law
for
the velocity cascade \cite{Kol,McComb}. However, analytical calculations
immediately
lead to difficulties: the simplest closure scheme to deal with the non linear
term in
the Navier-Stokes equation (`DIA') does not reproduce Kolmogorov's scaling
\cite{Kraich,McComb,Weich}. A lot of effort has been devoted, in particular by
Kraichnan, to understand why this was so. More refined schemes were proposed to
recover the $k^{-5/3}$ law and even to calculate adimensional constants
\cite{LDIA,RG,McComb}, but they are based on uncontrolled assumptions (like the
RG approach), or lead to enormous calculations, which are difficult to
manipulate and
are not fully transparent from a physical point of view \cite{LDIA}.

More recently, a tremendous activity has developed on the intermittency
 (or multifractal) corrections to Kolmogorov's scaling: higher moments of the
velocity
field do not seem to scale with the one originally predicted by Kolmogorov.
Many
interesting suggestions have been put forward to describe and explain this
feature, starting by Kolomgorov
himself \cite{Kol62,Mandel,Frish,Benzi,Arn,Cast,She,Dubru}.

In recent years, non linear partial differential equations with noise have been
the
 focus of quite a number of studies in the context of growing interfaces, with
the
`KPZ' equation standing out as a new paradigm \cite{FNS,KPZ,Medina,KPZRev}.
This equation is in
fact a variant of Burgers' equation, and has (together with its  many siblings)
a wide
range of applications in different physical contexts \cite{KPZRev}.
Interestingly,
this field is also related (through an appropriate `Cole-Hopf' mapping) to the
physics
of disordered systems, in particular elastic strings in random media (`directed
polymers'). The Burgers problem in $N$ dimensions is then equivalent to the
problem of a directed polymer in $N+1$ dimensions, which is the space time of
the original problem. The forcing term in the Burgers equation translates into
a random potential for the directed polymer, which is quenched in space time:
to each realization of the stirring force in the Burgers language corresponds
one sample of a directed polymer.
 This enables one to adapt techniques originally devised for with
spin-glasses and obtain original results on the underlying non-linear equation
(or
vice-versa) \cite{HHF,Kardar,MP,M,P,FH}. It is the aim of the present paper to
exploit
in detail this mapping, which allows us to propose an original {\it Ansatz} for
the velocity field for the
 randomly stirred
Burgers equation, which should become {\it exact} solution in high dimensions
and in the limit of large Reynolds number.
Scaling in the inertial range can then be
precisely discussed: we obtain in closed form the full probability distribution
for
velocity differences. In particular, the third moment of the velocity
difference grows
linearly with distance, i.e {\it \`a la} Kolmogorov. In fact, Kolmogorov's
dimensional
analysis should be directly applicable to Burgers' turbulence. We find however
very
strong intermittency effects, which we relate in a quantitative manner to the
existence of {\it large scale structures}, in the form of singularities
concentrated on time dependent
hypersurfaces.  Although not exact in finite dimension, we expect that our
(variational) description remains qualitatively correct even in one dimension.
We
also obtain the dynamical evolution of the velocity field: we find that the
field is
convected away by the largest structures, corresponding to a dynamical exponent
equal
to $z=1$ rather than the one obtained from Kolmogorov's scaling, $z_K=2/3$.

In the bulk of the paper, we primarily focus on the `dictionary' between
Burgers'
turbulence and disordered systems and discuss a number of physical points,
relegating
more technical points to various appendices. We show that Kolmogorov's scaling
has a
counterpart in the `directed polymer' language, where it is known as the
Larkin-Ovchinnikov scaling. It is however easy to see that this `naive' scaling
cannot
hold, as is confirmed by the full calculation. We describe in physical terms
the
nature of the velocity field, and argue that there should be a large distance
regime
(beyond the `injection' scale) characterized by a non-trivial exponent. We
briefly discuss
the problem of a passive scalar in such a velocity field. In the conclusion, we
compare our results to other approximation schemes, and comment on the possible
differences with `true' Navier-Stokes turbulence.

\section{ From Burgers turbulence to directed polymers: dictionary and
dimensional analysis}

The problem we shall consider is that of a randomly forced {\it potential}
 flow in $N$ dimensions, which follows the Burgers equation
{\footnote{The density of the fluid is taken equal to one.}}:
$$
{\partial \v \over \partial t} + (\v \cdot \vec \nabla) \ \v
= \nu \nabla^2 \v + \vec f(\vec x,t) \eqno(2.1)
$$
where $\v$ is minus the gradient of a velocity potential $\h$, and
$\vec f$ is a randomly fluctuating force, which is also minus the
gradient of a potential $\p$. $\v$ and $\vec x$ are $N$-dimensional
 vectors,
 and there is no constraint on $\vec \nabla . \v$.
Here, we wish to describe the problem of a fluid which
is randomly stirred {\it only at very large lengthscales}. We shall
thus take $\phi$ to be Gaussian, with fluctuations given by:
$$
{\overline{\p \phi(\vec x',t')}}
= {\epsilon \Delta^2 N}
\delta(t-t') \exp-\left[{(\vx-\vx')^2 \over 2 N \Delta^2}\right].\eqno(2.2)
 $$
Here $\Delta$ is the `injection' length, over which the forcing is
roughly constant. The correlation {\it time} of the forcing was set
to zero; a finite (small) correlation time would however not affect
 the following conclusions.
The stirring force correlation thus reads:
$$
\overline{f^\mu(\vec x,\tau) f^\nu(\vec x',\tau')} =
\epsilon \delta(\tau-\tau') \left[\delta^{\mu \nu} - {(\vec x - \vec x')^\mu
(\vec x - \vec x')^\nu \over N\Delta^2} \right] \
 \exp-\left[{(\vx-\vx')^2 \over 2 N \Delta^2}\right].\eqno(2.3)
$$
As will be clear below, this ensures that the injected energy per unit time,
defined as the increase of kinetic energy ${1 \over 2} \v^2$, is equal to
${N \over 2} \epsilon$ (Note that the dimension  of $\epsilon$ is
$[x^2]/[t^3]$).
 The
dependance of the force
correlations on the dimension $N$ has been chosen such as to insure the
existence of the large $N$ limit.

The typical velocity at the
injection scale is, from dimensional considerations,
$v_\Delta \equiv (\epsilon \Delta)^{1\over 3}$,
 which allows us to define the Reynolds number as:
$$
Re  \equiv {v_\Delta \Delta \over \nu} = ({\epsilon \Delta^4 \over
\nu^3})^{1\over 3} \ .\eqno(2.4)
$$
We shall be interested in studying the statistics of the velocity
field at large $Re$.

Let us now use the standard techniques to transform this problem into
a directed polymer.
 Integrating once Eq. (1), with $\v= - \vec \nabla h$ one finds the so-called
Kardar-Parisi-Zhang (KPZ) equation:
$$
{\partial \h \over \partial t} = {1\over 2} (\nabla \h)^2
+ \nu \nabla^2 \h + \p    \eqno(2.5)
$$
which describes, in particular, the growth of a surface under a
random `rain' of particles, the flux of which is given by $\p $. At
 this stage, the crucial difference with previous work on the KPZ
equation is in the correlations of the
noise $\p $. We shall thus be mostly concerned, in the following,
 with the velocity
 field statistics at length scales {\it smaller} than $\Delta$; for
length scales larger than $\Delta$, the KPZ scaling prevails.
Now, as is well known, the KPZ equation (2.5) can be transformed
 (through a `Hopf-Cole' transformation) into a linear problem
describing a directed line  (polymer) in a random potential. Setting
$\h = 2\nu \log \Z$, one finds that $\Z$ obeys the equation
$$
{\partial \Z \over \partial t}
= \nu \nabla^2 \Z + {1 \over 2\nu} \p \Z\eqno(2.6)
$$
which is the equation  for the partition function of an
 elastic string in a random potential $\V={1 \over 2\nu} \p $,
 subject to the constraint that its end point is fixed at
 $(\vec x,t)$. Said differently, the solution of Eq. (2.3) can be
written as a path integral (sum over all configurations):
$$
\Z =  \int\rho(\vx_0) d \vx_0
 \int_{\vec x(t=0)=\vx_0}^{\vec x(t)=\vec x} d[\vec x(\tau)] \exp (-{\cal H}) \
{}.
\eqno(2.7)
$$
The hamiltonian ${\cal H}$ is  given by:
$$
{\cal H} =   \int_0^t d\tau  \[[ {c \over 2} \(( {d\vec x \over d\tau} \)) ^2 +
V(\vec x(\tau),\tau) \]].\eqno(2.8)
$$

The `dictionary' between the two problems is the following:
The temperature scale of the polymer has been chosen equal to one. Then
the elastic modulus of the polymer is
$$
c \equiv {1 \over 2\nu} \ .\eqno(2.9)
$$
The random potential seen by the polymer $V$ has a gaussian distribution,
with a second moment given by:
$$
\overline{V(\vx ,\tau) V(\vx',\tau')}=
{W N}
\delta(t-t') \exp-\left[{(\vx-\vx')^2 \over 2 N \Delta^2}\right].\eqno(2.10)
 $$
where the strength of potential fluctuations, $W$, is
related to the energy density $\epsilon$ through:
 $$
W \equiv {\epsilon \Delta^2 \over 4\nu^2}\ ,\eqno(2.11)
$$
and the length scale of fluctuations of the potential is
equal to $\Delta$, the lengthscale at which the Burgers fluid is stirred.
The probability distribution $\rho(\vx_0)$ of the initial
point of the polymer is related to the initial conditions of the
velocity through:
$$
\vec{v}(\vx,t=0)= -2 \nu \vec \nabla \rho (\vx) \ .\eqno(2.12)
$$

The non-linearity has disappeared from this formulation, and has
 been replaced by the famous problem of disordered systems, which
 is to average
the logarithm of a partition function (in order to calculate various
 moments of the velocity field $\v$). In the next section we shall
 deal with this problem
 using the replica trick and a variational method which becomes exact
 in the limit of very large dimensions $N$.

Before turning to this calculation, it is useful to give some kind of
qualitative (more or less dimensional) analysis of the directed polymer,
and to enrich our  `dictionary' by stating its
 counterpart in the  turbulence
 language.
The study of an elastic structure like the directed polymer
 in presence of a random potential
has been discussed in several works recently \cite{DP,MP}. In the phase
where the disorder is strong, one expects
a scaling behaviour of the lateral fluctuations of the polymer
described by a wandering exponent $\zeta$:
$$
\overline{<\(( \vx(t) -\vx(t') \)) ^2 >} \simeq A |t-t'|^{2 \zeta}.\eqno(2.13)
$$
The thermal fluctuations are irrelevant at large distance and this scaling
also holds in the zero temperature limit, or for the disconnected
correlation $\overline{\(( <\vx(t) -\vx(t')> \)) ^2} $. As for the free energy
differences for two polymers finishing at point $t,\vx$ and $t',\vx'$, they
scale as:
$$
\overline{\(( h(\vx,t) -h(\vx',t') \)) ^2 } \simeq  |\vx -\vx'|^{2
\omega/\zeta}
g_h\(( {|\vx -\vx'| \over |t-t'|^{\zeta} }\)).\eqno(2.14)
$$
The $t\to t'$ limit then implies that the  free energy difference
at points $\vx$ and $\vx'$ scales as $|\vx-\vx'|^{2\omega/\zeta}$, and
therefore the difference of velocities in Burgers equation should scale
as:
$$
\overline{\(( \v(\vx,t)-\v(\vx ', t')\))^2} \simeq
|\vx(t) -\vx(t')| ^{(2 {\omega \over \zeta} -2)}
g_v\(( {|\vx(t) -\vx(t')| \over |t-t'|^{\zeta} }\)).\eqno(2.15)
$$
If this last scaling form holds, the galilean invariance of Burgers equation
implies that the two terms in  $\partial \v / \partial t $ and
$ (\v \cdot \vec \nabla) \ \v $ should scale in the same way under a rescaling
$ \vx \to b \vx$ and  $t \to b^{1/\zeta} t$, which implies $\omega=2 \zeta-1$
\cite{HHF,Medina}.
(Another way to argue about this identity directly
on the polymer problem is by observing that the
fluctuations of elastic energy in the directed polymer scale as
$x^2/t=t^{(2 \zeta-1)}=t^\omega$). So we are left with only one scaling
exponent.

An important point which was discussed in \cite{BMY,LeDG} is the existence of
two
distinct scaling regimes, in the case where the random potential has
 a large correlation length, $\Delta$. The regime which is most studied for
directed
polymer is the large time regime where the transverse fluctuations
of the polymer are much larger than $\Delta$. But  there  also
exists a short distance regime
 where the transverse
fluctuations
of the polymer are smaller than the correlation length of the potential.
This is obviously the regime which interests us most for the turbulence
problem (although we shall return to the long distance regime later on).
This regime holds for time differences shorter than a typical time scale
$\tau^*$, defined from:
$$
\overline{<\(( \vx(t+ \tau^*) -\vx(t) \)) ^2 >} \simeq \Delta^2  \
.\eqno(2.16)$$
One expects to be allowed to linearize the random potential in this slowly
varying
 regime, and to
study the much simpler (linear) problem of a polymer with a random force,
 defined by the Hamiltonian:
$$
{\cal H} =   \int_0^t d\tau  \[[ {c \over 2} \(( {d\vec x \over d\tau} \)) ^2 -
\vf_L(\tau) \cdot \vx(\tau) \]],\eqno(2.17)
$$
(where the random force is of order $|f_L| \simeq \sqrt{W}/\Delta$).
This random force problem was studied by Larkin and Ovchinikov long time ago
 \cite{LO}.
In this regime there is no metastable state and the problem is easily solved
for one
given sample. Assuming for simplicity periodic boundary conditions, one obtains
the Fourier transform of the average polymer's position::
$$
<\vx(\omega)>={\vf(\omega) \over c \ \omega^2}.
\eqno(2.18)
$$
A dimensional analysis of the Fourier transform would lead to:
$$
|<\vx(t)-\vx(t')>|^2 \simeq {W \over c^2\Delta^2} |t-t'|^3,
\eqno(2.19)
$$
from which one deduces the scaling exponent in this regime,
and the value of $\tau^*$:
$$
\zeta_K
 = {3 \over 2} \ ,
 \ \ \ \  \omega_K=
2 \zeta_K-1
=2 \ , \ \ \ \ \tau^* \equiv \((
{c^2 \Delta^4 \over W} \))^{1/3} \equiv {\Delta \over v_\Delta}\eqno(2.20)
$$
This result looks very nice when translated in terms of turbulence,
since it predicts (using (2.15)) that the
difference of velocity between two points at distance $r$ will scale
like $r^{\omega_K/\zeta_K -1} = r^{1/3}$, and also that time scales and length
scales
 are related through $t^{\zeta_K} \sim r$, with $\zeta_K=3/2$ (Richardson
diffusion).
Note that
$\tau^*$ is simply the convective time across the injection length $\Delta$.

These results thus precisely reproduce the Kolmogorov
scaling, which is here derived from a very simple argument on the
directed polymer problem! However, it turns out that this result is wrong,
because we must go beyond the linear approximation.

Technically the reason is in the integration which leads from the expression
(2.18) of the average position in Fourier space to the scaling expression
(2.19). Clearly the corresponding integral over frequencies $\omega$ is
divergent
at small $\omega$, and it turns out that the result is proportional
to $(t-t')^2 {\cal T}$,where $\cal T$ is the total length of the polymer.
One sees that, even if $t-t'<\tau^*$, the scaling depends on time scales
which are  larger than $\Delta$. Therefore one cannot work out the
scaling behaviour at small length and time scales within the linear --
random force -- approximation. In the polymer problem one must study the case
of a random potential, which is a non linear problem with many
metastable states \cite{MP,BMY}. Translated into the turbulence language, these
correspond to intermittency effects which are crucial and cannot be neglected.
The full solution derived in the
next section indeed
finds a Kolmogorov scaling, but only for the third moment of the
velocity difference $ (\v(\vx,t)-\v(\vec 0,t))^3 \simeq |\vx|$.
The other moments do not agree since all the moments
$ (\v(\vx,t)-\v(\vec 0,t))^q$ with
$q$ larger or equal to 1 scale like $|\vx|$. Furthermore,
 length and time are related by a
{\it convective} scaling, corresponding to the dynamical exponent $z={1\over
\zeta}=1$.

\section{
The replica variational approach}
\subsection{The replica solution}
We now turn to the problem of computing the average `free energy' $\overline
{\log \Z}$
of the directed polymer
 over the random forcing $V(\vx,\tau)$. The procedure we use is standard, and
we
follow closely \cite{MP,BMY}: we first express $\log \Z$ as the `zero replica'
limit
$\log Z = \lim_{n \to 0} {Z^n -1 \over n}$, average $Z^n$ which generates an
effective
attraction between replicas, which we treat using a Gaussian variational
Ansatz.
The quality of this approximation, and the regimes of dimension and Reynolds
number
 in which it becomes exact,
will be discussed in section (3.3).

We proceed as usual by writing the average of $Z^n$ as the partition function
of
the $n-$ replica Hamiltonian ${\cal H}_n$, which reads:
$$
{\cal H}_n = {1 \over 2} \sum_{a=1}^n  \int_0^t d\tau
\left[c  \(( {d\vec x_a \over d\tau} \))^2 +
 \mu     \vec x_a  ^2 \right]
$$
$$
-{W N \over 2} \sum_{a,b} \int_0^t d\tau
 \exp \left[-{(\vx_a(\tau)-\vx_b(\tau))^2 \over 2N\Delta^2}\right]\eqno(3.1)
$$
where we have added a `mass' term $\mu$ for regularizing intermediate
computations,
which we will set
eventually to zero. We furthermore consider periodic polymers, for which
 $\vec x(t)=\vec x(0)$. This is not exactly the same problem as the one with
free
ends which is in direct correspondence with Burgers's equation. We shall first
work
out the periodic case, and in section (3.3) we discuss the changes for the free
case.

In order to handle the problems of the metastable states,
the idea is to use a variational method and approximate ${\cal H}_n$ by an
effective
 Gaussian Hamiltonian
${\cal H}_v$, which we write in Fourier space  as:
$$
{\cal H}_v \equiv {1 \over 2} \sum_{a,b}  \dw  \ \vx_a(-\omega)
G^{-1}_{ab}(\omega)
\vx_b(\omega)\eqno(3.2)
$$
where $\dw$ stands for $\int_{-\infty}^{+\infty} {d\omega \over 2\pi}$.
Note that ${\cal H}_v$ is isotropic in real space; however its structure in
`replica'
space is arbitrary. The trial free-energy obtained with ${\cal H}_v$ depends on
$G_{ab}$ and reads:
 $$ {\cal F}_v[G] = <{\cal H}_n>_v - {N \over 2} Tr \log G
\eqno(3.3) $$
where $<...>_v$ means averaging with the Bolzmann weight associated to the
trial Hamiltonian ${\cal H}_v$. Remember that
 the temperature scale has been set to one.
The calculation of $ <{\cal H}_n>_v$ using Gaussian integrals is
straightforward and
leads to:
$$
 <{\cal H}_n>_v= {N {\cal T}\over 2} \(( \sum_{a} \dw (c\omega^2 + \mu) \Ga
- {W} \sum_{ab} \left(1+{B_{ab} \over N\Delta^2}\right)^{-{N\over 2}} \))
\eqno(3.4)
$$
where $\cal T$ is the total length of the polymer and
$B_{ab} \equiv \dw [\Ga+\Gb-2\Gab]$. The variational statement is that
${\cal F}_v[G]$ is greater than or equal to the `true' free-energy. We thus
look at
the optimal $G_{ab}$ such that ${\partial {\cal F}_v[G] \over \partial \Gab}
\equiv
0$. This leads to the following self-consistent equations:
$$
[G^{-1}]_{ab}(\omega) = - {W \over \Delta^2}
\left(1+{B_{ab} \over N\Delta^2}\right)^{-{N\over 2}-1} \quad (a \neq
b)\eqno(3.5)
$$
and
$$
\Ga + \sum_{b \neq a} \Gab \equiv \Gc = {1 \over \mu + c\omega^2}\eqno(3.6)
$$
The task is now to solve these equations using some Ansatz on the structure of
$G_{ab}$. This has been discussed in full details in \cite{MP}.
We keep here to dimensions $N > 2$. There are two regimes of Reynolds number,
separated by a critical value $Re_c=[2(1-2/N)^{(1-N/2)}]^{1/3}$. For $Re<Re_c$,
the solution is a 'replica symmetric' one with $\Gab=\delta_{ab} G_c(\omega)+
G(\omega)$. The two propagators $G_c$ and $G$ are easily computed. Translated
in terms
of the velocity, this just corresponds, for an infinite
size system (but for a finite injection length $\Delta$) to a vanishing
velocity field.
 For a finite `box' of length $L = \sqrt{\nu \over \mu}$, one finds that $\v
\simeq  \nu {(\vx -\vx_0) \over L^2}$, where $\vx_0$ is a random (Gaussian)
time
dependent variable depending on the forcing history.
When one increases the forcing beyond $Re_c$
the correct solution is the so-called `one step' replica symmetry
breaking scheme. This amounts to parametrizing the off-diagonal elements $\Gab$
with
{\it two} functions $\GG,\G1$, depending on whether the replica indices belong
to the
same `block' or to different blocks. The size of these blocks are furthermore
parametrized by a number $m$ ranging between 0 and 1, fixed by imposing that
${\partial {\cal F}_v[G_0,G_1,m] \over \partial m} = 0$. We refer the reader to
\cite{MPV,MP} for a more detailed discussion of this construction, but will
recall
below its physical interpretation, on which we shall heavily rely to discuss
our
results in the turbulence language. The relevant formulae for inverting such
matrices
are given, for completeness, in Appendix A, together with the basic calculation
steps
needed to solve Eqs. (3.5-6).

One important final equation is the one fixing $m$. We find that:
$$
(1-{2m \over N})^{1-{N/2}} = 2 m^3 {\Delta^2 W \over \nu} = {1\over 2} m^3 Re^3
\eqno(3.7)
$$
where we have used the definition of the Reynolds number (Eq. (2.4)).
Note that Eq. (3.7) is well behaved in the limit $N \to \infty$, where it
becomes
$2 e^m = (m Re)^3$.
We shall see later that when $Re > Re_c$, the breaking of replica symmetry
 corresponds to a non trivial structuration of the flow
into large size structures which are`cells' of size $\Delta$, inside which the
velocity
is of order
 $v_\Delta \sim Re {\nu \over \Delta}$ (independently of the box size $L$). We
shall
thus
identify $Re_c$ as the critical Reynolds number for the onset of turbulence.

\subsection{Physical description of the solution}
Let us see how this is encoded in the one-step solution, by first looking at
{\it equal time correlations}. Within the Gaussian variational Ansatz, one
finds that
$$
{\cal P}[\{\vx_a(t)\}] = Z^{-1}\sum_{\pi} \exp \(( -{1\over 2} \sum_{a,b}
[Q^{-1}]_{\pi(a),\pi(b)} \vx_a(t)
\vx_b(t) \)) \eqno(3.8)
$$
where $Q_{ab}=\dw G_{ab}(\omega)$, and  $\sum_{\pi}$ denotes the sum over
all the $n!$ permutations of replica indices. Taking the
limit where the mass term $\mu$ is zero, we find that (see Appendix A):
$$
Q_0 \sim \mu^{N/4-1} \to 0, \quad Q_1-Q_0 \to \infty, \quad
\tilde Q - Q_1 = {m \Delta^2 \over (1 - {2m \over N})}
\eqno(3.9) $$

In the very large Reynolds limit, on which we focus now, the parameter $m$
scales
as $1/Re$ and one finds
$\tilde Q - Q_1 \simeq  m {\Delta^2}$.

The physical interpretation of the replica probability distribution Eq. (3.8)
has been worked out in \cite{MP}, and is particularly simple in the present
case where
$Q_0$ is zero and $Q_1-Q_0$ is very large. Remember that one is dealing with a
polymer
in a random environment. For each `sample' $\Omega$ (i.e for a particular
realisation of
the disorder $V(\vec x,t)$) the probability distribution $P_\Omega(\vx)$ for
the `end
point' of the polymer $\vx$ has a certain shape. The 1 step replica
symmetry breaking variational approach
assumes that this distribution can be written as a weighted sum of Gaussians:
 $$
P_\Omega(\vx) = \sum_\a W_\a \exp \((-{1\over 2 \delta}[\vx-\vra]^2 \))
\eqno(3.10)
 $$
where
$\delta \equiv \tilde Q - Q_1$, $W_\a$ are random weights chosen with a
probability
proportional to $ W^{-1-m}(1-W)^{m-1}$ and $\vra$ are uniformly distributed in
the `box' of
size $L$
(A more formal description of the solution is presented in Appendix B, together
with useful technical details).
 The $W_\a$ and the $\vra$ encode the particular features of the sample
$\Omega$, and must thus eventually be averaged over.

The above construction was restricted to a certain `time' (or length of the
polymer).
Similar considerations also enable us to construct two-time correlation
functions, and
the result is very simple: it amounts to let the $\vra$ acquire a time
dependence.
More precisely, $\vra$ are independent Gaussian, time dependent variables such
that
(again taking the $\mu=0$ limit):
$$ C(t-t') \equiv \overline{<[\vra(t)-\vra(t')]^2>} =
{2 S \over m }\dw {(1-\cos \omega(t-t'))\over c\omega^2(c\omega^2 + {S}) } \ ,
\eqno(3.11)
$$
where $S=1/(4 c (\tilde Q- Q_1)^2)$.
Eq. (3.11) shows that a characteristic time scale appears, given by $
\sqrt{c\over S} = 2^{1/3} {\Delta^2 \over \nu Re}$, which is nothing but the
`Larkin-Ovchinnikov' time $\tau^*$ encountered above and corresponds to the
convective
time across the injection length. For $\tau \gg \tau^*$, one finds that
$C(\tau)=
2^{4\over 3} \Delta^2 {\tau \over \tau^*}$, whereas for $\tau \ll \tau^*$, one
finds
$C(\tau)= {1\over 2}  \Delta^2  ({\tau \over \tau^*})^2$.

Eqs. (3.10-11), together with the value of $m$ and $\delta$, are the central
results
of  this paper, from which we shall derive in the next section
the very interesting statistical properties
of the velocity field, in particular the exact calculation of the full
probability
distribution of $\v(\vx,t)-\vec v(\vx',t)$. Before turning to this computation,
we
first discuss the validity of this replica solution.

\subsection{Discussion of the replica solution}
 One is not able to solve exactly the directed polymer problem. The variational
method
to which we have resorted is however known to be a good approximation to the
real behaviour
of the directed polymer (\cite{MP,MPtoy}). An
interesting advantage of this approach is that it gives
exact answers for the thermodynamics in
the limit of a
large number of dimensions $N \to \infty$. Technically, as shown in \cite{MP},
this
comes from the fact that this variational Ansatz can be seen as a resummation
of the
`Hartree' diagrams in the perturbation theory for the correlation function,
 which are the only ones surviving in the
large $N$ limit.
 However beyond the thermodynamic potentials,
there are some quantities for which this approach fails
to give the right answer, even for $N \to \infty$.
It is unfortunately the case of the velocity correlation function
defined in (2.15), for which our calculation may only become exact in the large
$N$ {\it
and} large Reynolds limit.
The problem can be understood as
follows: the replica calculation is expected to reproduce faithfully the
thermodynamical behaviour of the system, by correctly describing the {\it
low-lying}
energy states. The weight of these low-lying states is found to be distributed
as $\propto W^{-1-m}(1-W)^{m-1}$. Hence the moments $\overline
{[P_\Omega(\vec
x)]^\eta}$ which are primarily determined by the low-lying states (i.e. by
those
with large weights) are expected to be accurate, while those which are
sensitive to the $W \to 0$ part of the distribution are not. This is the case
when
$\eta < m$, in particular for $\overline{\log P_\Omega(\vec x)}$ which we need
in order to
calculate the velocity in the Burgers language. Hence, only when $m \to 0$,
i.e. at
infinite Reynolds number, can the method be reliably used to determine
exactly the
statistical properties of the velocity field. To summarize this discussion
we expect that this approach may
be exact only when $N \to \infty$ {\it and} $m \to 0$. This issue can be
studied from the solution by a study of the correlation identities, some
of which will be checked in section (4.3). In any case the picture which
emerges from the discussion in the next section is an appealing one which seems
to be
a good approximation even at small $N$.

Eventually, we want to discuss an important detail. The previous
replica calculation was performed, for simplicity, in the
case of a periodic polymer, i.e. one for which $x(t) \equiv x(0)$. From the
Burgers
equation point of view, however, the end point of the `polymer' must be left
free. As
argued by Nelson and Vinokur in a different context \cite{Nel},
the difference between points far `inside'
the chain and boundary points is the fact that the statistical weight of the
former
is given, in quantum mechanical language \cite{dGe}, by the product of {\it
two}
propagators, while the end point only needs one. If the ground state wave
function
of the corresponding quantum mechanical Hamiltonian decays exponentially on a
characteristic length scale
$\ell$, this means that the lateral fluctuations of the polymer's end point
decay on length
scale $\ell$, while the fluctuations for points in the bulk decay on a length
scale
$\ell/2$. The study of the system with periodic boundary conditions which we
performed in Sect.(3.1) only deals with these fluctuations in the bulk.
The simple consequence of this analysis for
our problem is that formula (3.10) can be used for free end points as well,
although
the value of $\delta$ must be doubled, i.e.:
$$
\delta = 2 m \Delta^2 \ .
\eqno(3.12)
$$

\section{
Structure of the velocity field}

\subsection{Qualitative Arguments}

We can now use our `dictionary' between the two problems, in particular the
fact that the velocity field is given  by the derivative of the
free-energy of the polymer problem with respect to $\vx$. For a given
realisation of
the forcing, this free-energy is simply the logarithm of the probability
distribution
$P_\Omega(\vx)$ given in Eq. (3.10); one can thus express $\v$ as:
$$
 \v={2 \nu \ov
\de}{\sum_\a W_\a (\vx-\vra) e^{-(\vx-\vra)^2/2 \de} \ov \sum_\a W_\a
e^{-(\vx-\vra)^2/2 \de}}  \ .
 \eqno(4.1)
$$
We shall study the statistical properties of this velocity field at large
Reynolds numbers.
We recall that from the previous analysis, we have $ \delta = 2 m \Delta^2
\simeq
2^{4/3} \Delta^2 /Re$.
Typical snapshots of such a velocity  field in $N=1$ or $N=2$ dimensions are
given in
Fig. 1 a,b. Very clearly, a cellular structure appears. Within each cell, the
velocity
field is radial $\v \sim {2 \nu \ov \de}  (\vx-\vra)$, with a rapid variation
(shocks)
across the boundaries of these cells.

This structure can be understood qualitatively from Eq. (4.1), using the
important property
that the weights $W_\a$ have a very broad distribution.
More precisely, we know that $W_\a=\exp(-f_\alpha)/\sum_\gamma
\exp(-f_\gamma)$, where
the $f_\alpha$ are independant random variables with an exponential
distribution
increasing as $\exp(mf)$.
For $m $ small, there is a very strong hierarchy between the smallest $f$
appearing in the sum. For instance the gap between the smallest $f_\alpha$ and
the
next scales as $1/m \propto Re$. So the low lying $f_\alpha$, which are the
only
terms contributing to the sum, behave at large $Re$ as $f_\alpha \simeq \hat
f_\alpha Re$,
and the velocity field in this regime takes the form:
$$
 \v \ \sim \ {2 \nu \ov \de}
\ {\sum_\a (\vx-\vra) \exp \((Re \left[
 -\hat f_\alpha-{(\vx-\vra)^2 \over 2^{7/3} \Delta^2} \right] \))
 \ov
\sum_\a
  \exp \((Re \left[
 -\hat f_\alpha-{(\vx-\vra)^2 \over 2^{7/3} \Delta^2} \right] \))
}
 \ .
 \eqno(4.2)
$$
For a given $\vx$ and a large $Re$, the major
contribution to the sum comes from the largest term. This dominant term
will suddenly switch from --say-- $\a$ to $\beta$ when the space dependent
Gaussian
factor compensates the weight difference, i.e. when ${(\vx^* - \vra)^2 \sim
(\vx^* -
\vec r_\beta)^2 + {\de \over m}}$. This leads to a typical size of cells of
order
$\sqrt{\de \over m} \propto \Delta$, i.e. the injection length itself. The
width $\ell$
of the shock separating two cells is obtained by writing that ${|\vx^* -\vra|
\ell
\over \de } \sim 1$, i.e. $\ell \propto {\Delta \over Re} \ll \Delta$, which
shows that
the notion of cells is indeed well defined at large Reynolds.  This
construction allows
one to guess the structure of the probability distribution of the velocity
difference
between nearby points. Let us present the argument in $N=1$ dimension (its
generalisation to higher dimensions is straightforward), and call $x$ and $x+r$
these two points. The probability that a cell wall is present within the
interval
$[x,x+r]$ is obviously given by $p={r \over \Delta}$. In the limit where
$p$ is small, the velocity difference $u = v(x+r)-v(x)$ is equal to ${2
\nu r \ov \de}$ with probability $1-p$, and of order ${2 \nu \Delta \ov \de}
\sim {\nu Re \over \Delta} = v_\Delta$ with probability $p$. Hence:
$$
 P_{r}
(u) \sim (1-p) \delta\left(u -  {2 \nu r \ov \de}\right) + p
{1\over v_\Delta} f\left({u \over v_\Delta}\right)
 \eqno(4.3)
$$
where $f(.)$
is a certain scaling function.  This is true in the limit $Re \to \infty$ where
the
internal structure of the shocks can be neglected. We will see below that
precisely
such a form is obtained from an exact calculation. Let us now estimate the
various
moments of $u$ using expression (4.2):
$$
\overline{u^q} \sim (1-p)
({v_\Delta r \ov \Delta})^q + p \ A_q  v_\Delta^q  \ ,
\eqno(4.4)
 $$
where $A_q =
\int du u^q f(u)$. In the limit where $\ell \ll r \ll \Delta$, one finds that
for {\it all} $q$ larger than $1$, one has
 $$
 \overline{u^q} = A_q
v_\Delta^q  {r \over \Delta} \ ,
 \eqno(4.5)
$$
 whereas for $q<1$, one finds
 $$
\overline{u^q} \propto v_\Delta^q  ({r \over \Delta})^q \ .
 \eqno(4.6)
$$

 The velocity field is thus strongly intermittent, with a `multifractal'
spectrum
given in Fig. 2. Note however that the $q=3$ moment scales {\`a la} Kolmogorov,
i.e as
in the real turbulence problem. The presence of large scale structures (shocks)
forming an $N-$ dimensional froth-like pattern (see Fig. 1-b) is responsible
for such
a strong intermittency.
Large scale structures in `true' turbulence are similarly
thought to be the origin of the experimentally observed intermittency, which is
however much milder (see Fig. 2). The deep
reason of this difference is probably related to the `dimension' of the large
scale singularities, which is $N-1$ in the present case and only $1$ for vortex
lines in hydrodynamical turbulence.

 This strong intermittency, due to the cell and shock structure,
has already been discussed in the one dimensional Burgers turbulence
in the decaying (unforced) case (\cite{Burg1d,Frisch2,Esipov}). To the best of
our knowledge
the forced case has not been studied. Furthermore we are able to derive the
exact form
of the probability distribution function of velocity differences, to which
we now proceed.

\subsection{Exact results at infinite Reynolds number}

Using the direct evaluation of the moments of the velocity difference from Eq.
(4.1),
one can establish, after rather long manipulations detailed in Appendix B, the
following form for the full distribution of longitudinal velocity differences.
Let us first start by
the equal time case, and write $u = v^1(\vec x) - v^1(\vec y)$, where the
component
`1' is along the $\vx - \vec y$ axis, and $l=  {y^1-x^1 \over \Delta \sqrt{2}}$
(The other components would be treated in a similar manner). The result
then reads:

$$
P_l(u)=\de\((u-l  {2 \nu \over \delta }
\sqrt{ {\de \over m } }\)) e^{-l^2/8}
\int_{-\infty}^{\infty} Dh   R(h,\ml)
$$
$$
+\ml \int_{-\infty}^{\infty} dh
\int_{-h-\ml/2}^{\infty}Dt \int_{h-\ml/2}^{\infty} Ds
\de \(( u +(t+s){2 \nu \over \delta } \sqrt{ {\de \over m } } \mbox{sign}(l)
\))
R(h,\ml)^2  \ ,
\eqno(4.7)
$$
where $R(h,l)=\left[e^{-hl/2} {\cal M}_0(h-l/2) +  e^{hl/2}
{\cal M}_0(-h-\l/2)\right]^{-1}$ (${\cal M}_0$ is related to the standard error
function
 -- see Appendix B), and  $Dx \equiv {dx \over \sqrt{2\pi}} e^{-u^2 \over 2}$.
 Note that Eq. (4.6) has precisely the structure which we guessed
using qualitative arguments (Eq. (4.3)). We have tested this formula
numerically by
generating a one-dimensional velocity field using eq. (4.1), for $Re = 100$,
and
directly computed the second moment $\overline{u^2}$ as a function of $\rho$,
which we
compare in Fig 3 with the exact formula for $Re \to \infty$ obtained by
integrating
(4.7). For ${\Delta \over Re} \ll \rho \ll \Delta$, corresponding to the
`inertial range', one finds
$\overline{u^2} = {16 Re^2 \nu^2 \over \sqrt{\pi} \Delta^2} \ {|x-y| \over
\Delta}
$, or in terms of the `traditional' energy spectrum $E(k) \equiv
k^{N-1} \langle \v(\vec k)\v(-\vec k) \rangle$, $E(k) \propto k^{-2}$ for all
$N$.
Note the saturation for distances much larger than the cell size $\Delta$, on
which we
shall comment later. We have also computed the weight of the $\delta$ peak and
compared it to the one obtained analytically, with good agreement again.

We have also extended our analysis to two interesting situations. First of all,
one
may consider the case of a finite Reynolds number. The calculation of the
second
moment $\overline{u^2}$ shows that a new length scale appears{\footnote{Note
that $\ell_d^{(2)}$ is the width of the shocks $\ell$
introduced above}}  $\ell_d^{(2)} = {\Delta
\over Re} $, separating the above {\it linear} regime of $\overline{u^2}$  at
larger distances, from the {\it regular}, quadratic behaviour at small length
scales:
 $$
\overline{u^2} ={\epsilon \over 2\nu} r^2; \qquad  r < \ell_d^{(2)} \eqno(4.8)
$$
which is the standard result (up to the numerical prefactor) obtained in
turbulence
for length scales smaller than the dissipation length (see e.g. \cite{Landau}).
Note
that the `dissipation length' $\ell_d$ depends on the moment of the velocity
one
wishes to calculate: this is another consequence of intermittency. Matching the
regular
behaviour (4.7) with (4.4) or (4.6) suggests that the $q^{\mbox th}$ order
dissipation
length scales as $\ell_d^{(q)} = \Delta Re^{q \over 2 -2q}$; only for $q=3$
does one
recover the usual Kolmogorov dissipation length $\ell_d^{(3)}={\Delta \over
Re^{3/4}}$.

 From our Ansatz for the velocity field (4.1), and the fact that $\vec
r_\alpha(t)$ evolve according to (3.11), one can see that the full velocity
correlation function has the following scaling form (in the limit ${\Delta
\over Re} \ll r \ll \Delta$):
$$
\overline{[\vec v(\vx+\vec r,t) - \vec v(\vec x ,t')]^2} =  r \ g\left({r \over
\sqrt{2C(t'-t)}}\right) \eqno(4.9)
$$
where $C(\tau)$ is defined in Eq. (3.11) and $g(.)$ is a certain scaling
function. We have obtained the precise form of $g$ in the case where $t$ and
$t'$ are well `inside' the polymer, but did not attempt to compute it for `end
points' -- where additional numerical factors would appear (see the discussion
at the end of section 3.3).

In the limit $t-t' \ll \tau^*$ (see Eq. (3.11)), $\sqrt{2C(t-t')}= {v_\Delta}
|t-t'|$: Eq. (4.9) then means that fluctuations travel in a ballistic way with
a
`velocity' fixed by the injection scale. For larger time scales, the dynamics
recovers a diffusive character, since $\sqrt{2C(t-t')} \propto \Delta
\sqrt{|t-t'|\over
\tau^*}$. Note however that the effective diffusion constant $\nu_{\mbox turb.}
={ \Delta^2 \over \tau^* } \propto Re \  \nu$ is enormously enhanced compared
to its
`bare' value $\nu$ -- as is the case for usual turbulence. From Eq. (4.9) in
the limit $r \to 0$, one finds that for coinciding points, the velocity
difference grows with
time as:
 $$\overline{[\vec v(\vx,t) - \vec v(\vx,t')]^2} \propto \sqrt{2C(t-t')}
\eqno(4.10)$$

\subsection{`Sum-Rules' and the large Reynolds limit}

There are a number of `sum-rules' that the correct solution of the Burgers
equation
should satisfy, allowing us to test our prediction for the structure of the
velocity
field, as given by Eq. (4.1). The first and most interesting one physically is
the
`energy conservation', i.e.
$$
{1 \over 2} \overline {\partial \vec v^2 \over \partial t} = 0 = - \overline
{\vec v
\cdot (\vec v \cdot \vec \nabla) \vec v} + \nu \ \overline {\vec v \cdot \Delta
\vec v} +
\overline {\vec v \cdot \vec f}\eqno(4.11)
$$
The last term of this expression is the injected energy, and is equal to $+{N
\over
2} \epsilon$. The two first terms represent the `dissipated' energy -- notice
that contrarily to the incompressible Navier-Stokes case, this dissipation  is
non local (i.e. it cannot be expressed in terms of velocity derivatives only).
We find (see Appendix C, Eq. C.11), that the total dissipation is given by
$- {N \over
2} \epsilon (1-m)$. Hence we find
that the energy conservation is indeed exactly satisfied in the limit $m \to 0$
($Re \to
\infty$) {\it at fixed} $\epsilon$ (i.e. for $\Delta \to \infty$ or $\nu \to
0$). Interestingly, in this limit, this sum rule is true independently of $N$.
Eq. (4.11) can also be interpreted slightly differently. Suppose that at $t=0$
the
forcing is switched off. The subsequent evolution of the energy density is then
given by the dissipative terms, i.e. $N \Delta^{2/3}{d \epsilon^{2/3} \over dt}
\simeq - {N \epsilon \over 2}$, leading to $\epsilon \sim {\Delta^2 \over
t^3}$, or
else for the velocity scale $v_\Delta(t) \sim {\Delta \over t}$.

We have also checked that $\overline {\partial (\vec v^2)^2 \over \partial t} =
0$ in
the limit $Re \to \infty$ for all $N$, while $\overline {\partial \nabla \cdot
\vec v
\over \partial t} = 0$ identically for all $N$ and $Re$. The calculation of
velocity correlations for non coinciding points, such as ${\partial \over
\partial t} \overline {\vec v(\vx)
\vec v(\vec y)}$, are much more intricate, and are currently under
investigation.

Hence these direct checks of certain (local) correlation identities suggest
that our Ansatz might be exact for large $Re$.

\section{Discussion}

There are quite a number of points worth discussing under the light of the
previous
results, before commenting on the similarities and differences with the `real'
turbulence problem.

\subsection{Burgers turbulence in small dimensions}

As stated above, our results are {\it a priori} only exact for large dimensions
and
Reynolds number, the method we have used becoming a variational approximation
in finite
dimensions. By comparison with other problems where this approach was used,
however,
one expects that our solution describes faithfully the physical situation, even
for
relatively small values of $N$. Furthermore, all the sum-rules which we have
checked so far (Section 4.3) are satisfied for {\it any} $N$ in the limit $Re
\to
\infty$. Hence, we believe that the forced Burgers equation in say $N=3$
dimensions
will produce a cellular arrangement of the flow pattern, as described above. It
would
be extremely interesting to test this prediction numerically, as well as to
measure the
intermittency corrections -- which might be weaker in finite dimensions than
the ones
obtained within our Ansatz.

For $N \leq 2$, one however knows from previous studies that the structure of
the replica solution changes: instead of a `one step' breaking, a full
continuous
breaking scheme is needed. This raises the interesting question of knowing
whether
such a scheme could describe a more complicated behaviour of the moments of the
velocity field than the one encountered above. We do not have a complete answer
to
this question, but it seems that the solution obtained for $N \leq 2$ does not
lead
to a qualitatively different picture, the reason being that all the length
scales
appearing in recursive construction of the sample dependent measure $P_\Omega$
\cite{MP} (i.e.
the width of the Gaussians), as well as the `$m$' parameters describing the
weight
distributions, are all of the same order of magnitude. Thus, forced Burgers
turbulence in one dimension is expected
to look very much like the picture shown above (Fig 1-a), as indeed is well
known from studies of `decaying' turbulence from random initial conditions
\cite{Burg1d,Frisch2,Esipov}.

More work on this aspect would certainly be interesting, in particular to study
a variant of the present problem for $N=1$, where $f$ -- rather that $\phi$ --
is a random noise (see \cite{HK,RBE}). In this case also, a full breaking of
replica symetry is needed to describe the velocity field.

\subsection{Long-distance scaling of the velocity field}

As discussed in Section 2, the main difference between this work and previous
studies
on the directed polymer problem lies in the regime of length scales. Usually,
in the
directed polymer problem, one assumes that the correlation length of the
potential
$\Delta$ is very small, and one is interested in the long-distance $x \gg
\Delta$ scaling behaviour of -- say, the free-energy, where non trivial
exponents
appear. On the other hand, in line with most studies of the turbulence problem,
we
focused on the contrary on the {\it small} scales $x \ll \Delta$ regime, where
an
inertial range appears, characterized by an energy cascade $E(k)\propto
k^{-2}$. The
existence of these two regimes for the Burgers equation suggests that a similar
situation may also occur in `real' turbulence, where a non trivial scaling
regime
could exist for $x \gg \Delta$, characterized by a velocity correlation
converging
towards its asymptotic value as a power law:
$$
\overline{ [\v(\vx + \vec r)-\v(\vx)]^2} \simeq 2 \overline{ \v(\vx)^2}
 - { B \over
r^{2(1-{\omega \over \zeta})}} + ... \qquad r \gg \Delta \eqno(5.1)
$$
or equivalently, by a new exponent for the energy spectrum:
$E(k) \propto k^{1-2
{\omega \over \zeta}}$, where $\omega$ and $\zeta$ are the generalisation of
the
exponents defined for the polymer problem in Eqs. (2.13-14), corresponding now
to the usual large scale situation $r \gg \Delta$ studied in
directed polymers. Numerical results on directed polymers in
3+1 dimensions \cite{KPZRev} give
$\zeta \sim 0.6 , \ \omega =2 \zeta-1 \sim 0.2$. It would be very interesting
to analyze experimental data beyond the injection length $\Delta$ along these
lines.

\subsection{Passive scalar dispersion}

A subject of recent debate is the behaviour of a passive scalar in turbulent
flows,
which also shows experimentally `anomalous' density fluctuations -- in
particular
exponential tails \cite{exptails}. Although a detailed
study is beyond the scope of the present paper, it
is interesting to discuss qualitatively this problem for Burgers' turbulence.
Let us
suppose that the diffusing scalar obeys a Langevin equation of the form
$$
{d \vx \over dt} = \v(\vx,t) + \vec \eta(t) \eqno(5.2)
$$
where $\v(\vx,t)$ obeys the forced Burgers equation Eq. (2.1), and $\eta(t)$
describes the molecular diffusion, with diffusion constant equal to $D$. In the
`adiabatic' limit where the evolution of the velocity field can be assumed to
be small
compared to the equilibration time (i.e $D \gg v_\Delta \Delta$), the tracer
concentration  $\theta(\vx,t)$ is simply given by the Boltzman equilibrium:
$$
\theta(\vx,t) \propto \exp -{h(\vx,t) \over D} \eqno(5.3)
$$
where $h(\vx,t)$ is the velocity potential introduced in Eq. (2.5). Using
$h(\vx,t)=
 2 \nu \log P_\Omega(\vx,t) + {\mbox {cst.}}$, one finds that:
$$
\theta(\vx,t)  \propto [P_{\Omega}(\vx,t)]^{-{2\nu \over D}}\eqno(5.4)
$$

Now the full distribution function of $P_\Omega$ can be exactly calculated, and
is found to be a totally asymmetric L\'evy distribution $L_\mu$ with index
$\mu=m$. This in turn allows one to obtain the full distribution of $\theta$,
as
$$
{\cal P} (\theta) = \theta^{-1-{D \over 2\nu}} L_m(\theta^{-{D \over 2\nu}})
\eqno(5.5)
$$
Hence, we find that the {\it tail} of the
tracer density distribution is a {\it power-law} for small $\theta$, ${\cal P}
(\theta) \sim \theta^{-1+{Dm \over
2\nu}}$, which becomes broader and broader as the Reynolds number increases ($m
\propto Re^{-1}$). For large $\theta$, ${\cal P}(\theta)\sim \theta^{{Dm \over
2\nu}-1} \exp\left [-{\mbox {cst}} \theta^{{Dm \over
2\nu}}\right]$. Note however that our adiabatic assumption breaks down
precisely
when $ {Dm \over \nu} < 1$. It would be interesting to understand the nature of
the fluctuations of $\theta$ in the large Reynolds regime.

Hence we have argued that the fluctuations of tracer concentration are strongly
anomalous in Burgers's turbulence. In more physical terms, Eq. (5.2) shows that
the
tracer is convected towards the shock regions, where the concentration piles
up.
High density `sheets' of particles spontaneously form, revealing the
`froth-like' structure of the flow. This is similar to what happens for the
Burgers
equation with random initial conditions (but no forcing) \cite{Verga}, and
might be
relevant for astrophysical applications \cite{Zeldo}.

Finally, it must be noted that the temporal dispersion of tracers is {\it
convective} at short times, since it essentially follows the velocity field
which evolves according to Eq.(3.11). This is, as noted above, in contrast with
Richardson's diffusion, which states that $x \simeq t^{3\over 2}$. Convective
dispersion is however rather commonly observed (see \cite{Okubo}).

\subsection{Turbulence versus `Burgulence'}

The differences between Burgers' turbulence (coined `Burgulence' by Fournier
and Frisch in \cite{Burg1d}) and
hydrodynamical turbulence have been discussed many times. The most important
one is
the absence of vortex line singularities in a potential flow, which are thought
to
play an important role in turbulence \cite{vortex}. In our case, singularities
are concentrated on $N-1$ structures, giving rise to much larger
`intermittency' corrections.

Second, the energy is only dissipated by
viscosity at small length scales in turbulence, while there is an additional
dissipation term in Burgulence (which is, as discussed above, non-local).
This might be another important difference although, as shown above, it does
not
prevent the existence of a well-defined inertial range where the energy
spectrum
follows a $k^{-2}$ decay.

A third difference invoked by Kraichnan is that in fluids,
the incompressibility condition is maintained by the underlying pressure field,
which
would play an important role in the dynamics of the fluctuations and be at the
origin
of the Richardson-Kolmogorov scaling  $x \sim t^{3/2}$, instead of the
convection law $x \sim
v_\Delta t$ found here. On the other hand, the Burgers and incompressible
Navier-Stokes equations look very similar, in particular from a dimensional
analysis
point of view, which is at the heart of Kolmogorov's argument; furthermore, the
third moment of the velocity difference indeed scale in the same way (at least
for large $N$).

On a technical level, the simplest closure scheme for turbulence is
Kraichnan's DIA, which leads {\it both} for the Navier-Stokes equation and for
the
Burgers' equation, to a convective dynamics $x \sim v_\Delta t$ and to a
$k^{-3/2}$
energy spectrum. However, DIA's extension to
Lagrangian coordinates (LDIA) was argued by Kraichnan \cite{LDIA} to reproduce
exactly Kolmogorov's
scaling for the Navier-Stokes equation, but not for the Burgers case, where
results
similar to some of ours are obtained \footnote{Although the situation
investigated by Kraichnan was that of `decaying' turbulence; the intermittent
corrections were furthermore not discussed}.  It would be interesting to
investigate
the precise relation between the two seemingly very different approaches, and
in
particular to compare LDIA to our  results.

\subsection{Conclusions}

We have thus used a method inspired from spin-glasses to investigate a toy
model of
turbulence, and to propose an  Ansatz
 for the structure of the velocity field, which should  become exact in high
dimensions and when the Reynolds number is also large. We find that beyond a
critical
Reynolds number, there exists a well defined inertial range where the energy
spectrum
decreases as $k^{-2}$. The scaling variable is $x \over v_\Delta t$, where
$v_\Delta$
is the velocity at the injection scale $\Delta$. The third moment of the
velocity
difference scales linearly with distance, {\`a la} Kolmogorov, but strong
intermittent
corrections come into play, due to the presence of shocks localized on a
froth-like,
cellular pattern. The full distribution for the velocity difference is obtained
exactly. Interesting scaling results are also argued to hold at scales {\it
larger}
than the injection scale, and we suggest that experimental data on grid
turbulence
could be analyzed accordingly. Our results are presumably qualitatively correct
in low
dimension; numerical simulations would be welcome. We have discussed
qualitatively the
passive scalar problem and we have found power-law tails in the concentration
distribution, reflecting the localisation of the tracers near the shocks.

 From a technical point of view, it would be interesting to understand the
precise
relation between the present approach and Kraichnan's Lagrangian DIA, which
gives
the same scaling as the ones obtained here. From a different point of view, one
could also generalize this work to the case of a space-correlated forcing term
$f(\vx,t)$ with a {\it power-law} correlation function. This is the starting
point
of the renormalisation group (RG) analysis of turbulence: the exponent
describing the
decay of the power-law function is chosen as to reproduce Kolmogorov's scaling
and
an RG procedure \cite{RG} is applied to obtain adimensional prefactors (the
Kolmogorov
constant). The model studied here could provide an interesting benchmark to
discuss
the validity of such a procedure. In fact it can be used to test
all the various approximate methods which
have been introduced in the study of fully developped turbulence.

Finally, our Ansatz for the velocity field [Eq(4.1)], which is inspired from
our `replica' approach to the problem, has some interesting mathematical
properties
(see in particular Appendices B and C).
 It could be fruitful to  generalize Eq. (4.1) to describe rotational flows.

{\bf{Acnowledgements}}

MM thanks the SPhT at the CEA Saclay for its kind hospitality.
 This work has been supported
by the EC grant CHRX CT92 0063.

\vskip 1cm

\centerline{\bf Appendix A: Solution of the saddle-point equations.}

\vskip 1cm

The starting point of the one-step replica symmetry breaking calculation is the
expression of the free-energy in terms of $\Gt,\GG,\G1$ and $m$. Using the
expression
(AII.11) given in \cite{MP} for the trace of the logarithm of a one-step Parisi
matrix, we find that:
$$
{\cal F}= {1\over 2} \dw (c\omega^2 + \mu) \Gt + {W \over 2} \{(1-m)(1+{B_1
\over
N\Delta^2})^{-N/2} + m (1+{B_0 \over
N\Delta^2})^{-N/2}\}
$$
$$-{1 \over 2} \dw \ \left[{ 1 \over m} \log (\Gt - m\GG - (1-m) \G1) \right.$$
$$
\left. +
{\GG \over \Gt - m\GG - (1-m) \G1} - {1-m \over m} \log (\Gt -\G1)
\right] \eqno(A.1)
$$
where $B_{i} \equiv 2\dw [\Gt-G_i(\omega)]$. (We have set the temperature to
$T=1$).
Differentating $\cal F$ with respect to $\Gt,\GG,\G1$ yields equations (3.5-6)
specialized to the one-step solution, with
$$\Gc \equiv \Gt -  m\GG - (1-m) \G1 = {1 \over [G^{-1}]_c}= {1 \over (\mu +
c\omega^2)}\eqno(A.2),$$
$$
\GG = -{[G^{-1}]_0 \over (\mu + c\omega^2)^2}\eqno(A.3)
$$
and
$$
\G1 - \GG = {1 \over (\mu + c\omega^2)} { S \over m(\mu + c\omega^2 + S)}
\eqno(A.4)
$$
where we have introduced $S \equiv m([G^{-1}]_0 - [G^{-1}]_1)$. Using the
definitions of $B_0,B_1$, we obtain the equations:
$$
B_1 - B_0 = -{2 \over m} \dw { 1 \over \mu + c\omega^2 } { S \over \mu +
c\omega^2 +S
}\eqno(A.5)
$$
and, using (A.2)
$$m B_0 + (1-m) B_1 = 2 \dw [\Gt -  m\GG - (1-m) \G1] = 2 \dw { 1 \over \mu +
c\omega^2 }\eqno(A.6)
$$
from which we can deduce $B_0$ and $B_1$ as a function of $S$. Inserting these
values
into Eq. (3.5) then leads, in the limit $\mu \to 0$, to:
$$
[G^{-1}]_0 \sim 0; \qquad S = - m [G^{-1}]_1 = {m W \over \Delta^2} [1 + {1
\over
N \Delta^2 \sqrt{Sc}}]^{-1 - N/2}\eqno(A.7)
$$
Differentiating now (A.1) with respect to $m$, after a few manipulations, leads
to
$$
\sqrt{S \over c} = 2 m^2 W [1 + {1 \over
N \Delta^2 \sqrt{Sc}}]^{- N/2}\eqno(A.8)
$$
Eqs (A.7-8) allows us to obtain both $S$ and $m$ as:
$$
4 m^3 W\Delta^2 c = (1 - {2m \over N})^{1-N/2}
\eqno(A.9)
$$
and
$$
S = {(1 - {2m \over N})^2 \over 4 m^2 \Delta^4 c }
\eqno(A.10)
$$
Using Eqs (A2-7), we thus obtain in the limit $\mu \to 0$:
$$Q_c = {1 \over 2 \sqrt{\mu c}}, \quad Q_1 - Q_0 = {1 \over 2m \sqrt{\mu c}},
\quad \tilde Q-Q_1 = {1 \over 2 \sqrt{S c}} \eqno(A.11)$$
and $Q_0 \propto \mu^{{N \over 4 }-1}$. From this, we also obtain the inverse
$R$  of the matrix $Q$ as:
$$R_c = {1 \over Q_c} = 2 \sqrt{\mu c}, \quad R_0 = -{Q_0 \over Q_c^2} \propto
\mu^{N\over 4}, $$
$$ R_1-R_0 = {Q_0-Q_1 \over Q_c[Q_c+m(Q_0-Q_1)]} = {2 \sqrt{Sc} \over m}
\eqno(A.12)
$$

\vfill
\eject

\def\dv{{\mbox{\boldmath{$\delta v$}}}}
\def\cP{\cal P}
\def \bE{\bf E}
\def\fa{f_\alpha}
\def\fb{f_\beta}
\def\ra{r_\alpha}
\def\rb{r_\beta}
\def\l{\lambda}
\def\m{\mu}
\def \v{ {\vec v}}
\def \vv{ {\vec v}}
\def \vx{\vec x}
\def \vy{\vec y}
\def \vk{\vec k}
\def \vf{\vec f}
\def \vr{\vec r}
\def \h{ {h(\vec x,t)}}
\def \p{ {\phi(\vec x,t)}}
\def \Z{ {Z(\vec x,t)}}
\def \V{ {V(\vec x,t)}}
\def \r{ {m(\vec x,t)}}
\def \f{ {\vec f(\vec k,t)}}
\def \e2{ {\eta(-\vec k,-\omega)}}
\def \dr{ {\delta m}}
\def \rp {{r_\perp}}
\def \d{ {\delta m(\vec x,t)}}
\def \tx{\tilde x}
\def \ty{\tilde y}
\def\cP{\cal P}
\def\cC{\cal C}
\def\cM{\cal M}
\def \bE{\bf E}
\def\fa{f_\alpha}
\def\fb{f_\beta}
\def\ra{{\vec r}_\alpha}
\def\rb{{\vec r}_\beta}
\def\rg{{\vec r}_\gamma}
\def\l{\lambda}
\def\m{\mu}
\def\g{\gamma}

{\bf Appendix B: Probability distribution of the velocity difference}

\vskip 1cm

In this Appendix we compute the probability distribution function (pdf) of the
difference of velocities
between two points $\vx$ and $\vy$, projected onto $\vx-\vy$,
 at large Reynolds numbers. We shall proceed in three steps:
 first we show how to compute
the first few moments for general Reynolds numbers. Then we show how the
expression simplifies for large Re. This simplification is such that
we can extract in this large Re limit all moments and deduce from
it the pdf.

Our starting point is the  result from the replica computation,
which provides the following random process to build up the
velocity distribution: at a given time, the velocity field is
given by:

$$
\vv(\vx)={2 \nu \ov \de}{\sum_\a W_\a (\vx-\ra) e^{-(\vx-\ra)^2/2 \de}
\ov
\sum_\a W_\a e^{-(\vx-\ra)^2/2 \de}}
\eqno(B.1)
$$

The sum over $\a$ goes from $1$ to $M$. The points $\ra$ are uniformly
distributed in the volume $[-L/2,L/2]^d$. The weights $W_\a$
are random numbers drawn as follows: First one chooses M ``energies"
$f_\a$ at random, they are identically distributed
independent random variables with a probability distribution:

$$
\cP(f)= m e^{m(f-f_c)} \theta(f_c-f)
\eqno(B.2)
$$

Then the weights are given by:
$$
W_\a= {e^{-f_\a} \ov \sum_\b e^{-f_\b}}
\eqno(B.3)
$$

We shall let $M$, $L$, and $f_c$ go to infinity together, keeping
the density of states $M e^{-m f_c}/L$ fixed. The $\ra$ and $f_\a$
are uncorrelated. We shall denote respectively
by $\bE_r$ and $\bE_f$ the expectation
values with respect to these two sets of random variables.

Let us first evaluate the second moment. We take $\vx-\vy$ along the first
axis,
and compute the longitudinal correlation:
$$
\overline{v_1(x) v_1(y)}= \(({2 \nu \ov \de}\))^2
$$ $$
\bE_{r,f} \((
{
\sum_{\a, \b} e^{-\fa-\fb} (\vx-\ra)_1 (\vy-\rb)_1
e^{-[(\vx-\ra)^2+(\vy-\rb)^2]/2 \de}
\ov
\sum_{\a, \b} e^{-\fa-\fb}
e^{-[(\vx-\ra)^2+(\vy-\rb)^2]/2 \de}
} \)).
\eqno(B.4)
$$
We shall compute separately the  two contributions to this correlation
which come from the terms $\a=\b$ and $\a \ne \b$ in the numerator of
(B.4). We
thus write:
$$
\overline{v_1(x) v_1(y)} \equiv \(({2 \nu \ov \de}\))^2 (g_{11}+g_{12}),
\eqno(B.5)
$$
where $g_{11}$ and $g_{12}$ respectively contain
the terms $\a=\b$ and $\a \ne \b$ in
(B.4).

To compute $g_{11}$ we write the denominator of (B.4) in integral form, which
gives:
$$
g_{11}=
M \bE_{r,f} \int_0^\infty d\l \int_0^\infty d\m
e^{-2 \fa} (x-\ra)_1 (y-\ra)_1 e^{-[(x-\ra)^2+(y-\ra)^2]/2 \de}
 $$ $$
\exp\[[e^{-\fa} \(( \l e^{-(x-\ra)^2/2 \de}+ \m
e^{-(y-\ra)^2/2 \de} \)) \]]
 $$   $$
\prod_{\b (\ne \a)} \((
\exp\[[e^{-\fb} \(( \l e^{-(x-\rb)^2/2 \de}+ \m
e^{-(y-\rb)^2/2 \de} \)) \]] \))
\eqno(B.6)
$$
As for the mean over the free energies, we shall use
repeatedly in this Appendix the following formula, valid for large $f_c$:
$$
\int_{-\infty}^{f_c} df m e^{m(f-f_c)} \ e^{-k f -A e^{-f}}=
\cases{
m e^{-m f_c} A^{m-k} \Gamma(k-m),&if $k\ge1$;\cr
1- e^{-m f_c} A^{m} \Gamma(1-m) ,&if $k=0$.\cr
}
\eqno(B.7)
$$
After taking this average over the free energie, (B.6) becomes:
$$
g_{11}=
M \bE_{r} \int_0^\infty d\l \int_0^\infty d\m \ \
(\vx-\ra)_1 (\vy-\ra)_1 e^{-[(\vx-\ra)^2+(\vy-\ra)^2]/2 \de}
 $$
 $$
 \ m e^{-m f_c} \Gamma(2-m)
\(( \l e^{-(\vx-\ra)^2/2 \de}+ \m
e^{-(\vy-\ra)^2/2 \de} \))^{m-2}
 $$
 $$
\prod_{\b (\ne \a)} \[[
1-e^{-m f_c} \Gamma(1-m)
\(( \l e^{-(\vx-\rb)^2/2 \de}+ \m
e^{-(\vy-\rb)^2/2 \de} \))^m \]]
\eqno(B.8)
$$

We now average over the values of $\ra$ and $\rb$, with a uniform measure
in a box of size $L^N$. In the limit of large $L,M ,f_c$ at
fixed density we get:

 $$
g_{11}=m \Gamma(2-m)
{M e^{-m f_c} \over L}
\int_0^\infty d\l \int_0^\infty d\m
 $$
 $$
\int d{\vec r}
(\vx-\vr)_1 (\vy-\vr)_1 e^{-[(\vx-\vr)^2+(\vy-\vr)^2]/2 \de}
 \(( \l e^{-(\vx-\vr)^2/2 \de}+ \m
e^{-(\vy-\vr)^2/2 \de} \))^{m-2}
 $$
 $$
\exp \[[-{M
e^{-m f_c} \over L} \Gamma(1-m)
\int d\vr
\(( \l e^{-(\vx-\vr)^2/2 \de}+ \m
e^{-(\vy-\vr)^2/2 \de} \))^m \]]
\eqno(B.9)
 $$
It is convenient to rewrite $\mu \to \lambda \m'$, and integrate over
$\lambda$, which gives:
$$
g_{11}=(1-m)
 \int_0^\infty d\mu
 $$
$$
\[[
{
\int dr
(x-r)(y-r)e^{-[(x-r)^2+(y-r)^2]/2 \de}
 \(( e^{-(x-r)^2/2 \de}+ \mu
e^{-(y-r)^2/2 \de} \))^{m-2}
\over
\int dr
\(( e^{-(x-r)^2/2 \de}+ \mu
e^{-(y-r)^2/2 \de} \))^m
} \ \]].
\eqno(B.10)
$$
(In this expression the $\int dr$ is over a single $r$ variable, namely
the component of $\vr$ in the direction of $\vx-\vy$, the other components
have already been integrated out, their contributions cancel between the
 numerator and the denominator. We denote  $x=\vx_1$, and similarly $y=\vy_1$)

It is straightforward to perform the same steps for the second
contribution $g_{12}$ in (B.5). One finds:
$$
g_{12}=m
 \int_0^\infty d\mu \ \
\[[
\int dr
(x-r) e^{-(x-r)^2/2 \de}
 \(( e^{-(x-r)^2/2 \de}+ \mu
e^{-(y-r)^2/2 \de} \))^{m-1} \]]
 $$ $$
\[[
\int dr
 (y-r) e^{-(y-r)^2/2 \de}
 \(( e^{-(x-r)^2/2 \de}+ \mu
e^{-(y-r)^2/2 \de} \))^{m-1} \]]
 $$ $$
\[[
\int dr
\(( e^{-(x-r)^2/2 \de}+ \mu
e^{-(y-r)^2/2 \de} \))^m \]]^{-2}
 \ .
\eqno(B.11)
$$
It should be clear that these techniques allow to write the
velocity correlations of low order in a relatively closed
form (this means reduced to some finite dimensional integrals,
where all the averages over $r$ and $f$ have been taken care
of, as well as the limits $M \to \infty$, $L \to \infty$ and
$f_c \to \infty$). However the expressions are complicated
enough, especially when one goes to high moments,
and we have not found a closed form for the pdf in general.

Fortunately the situation simplifies in the limit of large
Reynolds numbers. Let us first work out the expression of the second
moment in this $Re \to \infty$ limit. We know from the replica solution
that in this limit the width $\de$ scales as $\de \simeq 2 m \Delta^2$,
where $\Delta$ is the scale at which energy is injected, and
$m$, the breakpoint in Parisi's RSB solution, behaves as
$m \simeq 2^{1/3}/Re$. Hereafter we shall use $m$ instead of $Re$,
and we want to understand the small $m$ limit of the velocity pdf.

We proceed and first work out the small $m$ limit of the two
point correlation. In $g_{11}$ we change variable to $x=\tx \sqrt{\de/m}$,
$y=\ty \sqrt{\de /m}$; We shall take the $m \to 0$ (large $Re$) limit
keeping $\tx$, and $\ty$ fixed. The algebraic distance
between the two points  is measured by
$l=(\tx-\ty)= (\vx-\vy)_1/(\Delta \sqrt{2})$. We also change the dummy
integration
variables in (B.10) from $r$ to $z=r \sqrt{\de/m} -(\tx+\ty)/2$,
and from $\m$ to $h=-m \ln(\m)/|l|$. This gives:
$$
g_{11} =
(1-m) \(( {\de \over m^2} \)) |l| \int_{-\infty}^\infty
 dh e^{-h |l| /m}
{I_{1,1}(h)
\over
I_{0,0}(h) \ ,
}
\eqno(B.12)
$$
where we have introduced the functions (defined for
integer $k_1,k_2$):

$$
I_{k_1,k_2}(h) \equiv \int {dz \over \sqrt{2\pi}}
(l/2-z)^{k_1} (-l/2-z)^{k_2}  \ \ \ \
 $$
$$
 e^{-z^2/2+l z (k_1-k_2)/2m}
\((e^{lz/2m} + e^{-h |l| /m-lz/2m}\))^{m-(k_1+k_2)}.
\eqno(B.13)
$$
Their small $m$ limit is easily worked out by a saddle point
integration. The result
for $k_1 \ge 1,k_2 \ge 1$ is:
$$
I_{k_1,k_2}(h) \sim_{m\to 0} \  {m \over |l|} \mbox{sign}(l)^{k_1+k_2}
\ e^{h \ml (k_2/m-1/2)} \
{e^{-h^2/2} \over \sqrt{2 \pi}}
\ \ \ \  $$ $$
\(( {\ml \over 2}+h\))^{k_1} \(( -{\ml \over 2}+h\))^{k_2}
{\Gamma(k_1)\Gamma(k_2) \over \Gamma(k_1+k_2)}
\eqno(B.14)
$$
While the results for $k_1 k_2=0$ read:
$$
I_{k_1,0}(h) \sim_{m\to 0 } \ e^{l^2/8} \  \mbox{sign}(l)^{k_1} \ \cM_{k_1}
\((-h-{\ml \over 2}\)),
$$ $$
I_{0,k_2}(h) \sim_{m\to 0 } \ e^{l^2/8} \  \mbox{sign}(l)^{k_2} \ \cM_{k_1}
\((h-{\ml \over 2}\)) e^{h \ml (k_2/m-1)},
$$ $$
I_{0,0}(h) \sim_{m\to 0 } \  e^{l^2/8-h \ml/2} \ \[[
e^{-h \ml/2} \cM_0(h-{\ml \over 2}) + e^{h \ml/2} \cM_0(-h-{\ml \over 2}) \]].
\eqno(B.15)
$$
In these equations we have used the definition:
$$
\cM_k(x) \equiv \int_x^\infty {dz \over \sqrt{2\pi}} z^k e^{-z^2/2}.
\eqno(B.16)
$$

Therefore we get for the $m\to 0$ limit of $g_{11}$:
$$
g_{11}={\de \over m} \sqrt{{2\over \pi}} \int_0^\infty dh
e^{-h^2/2} \ \ \ \
$$ $$
{ h^2-l^2/4 \over
 e^{l^2/8} \[[
e^{-h \ml/2} \cM_0(h-{\ml \over 2}) + e^{h \ml/2} \cM_0(-h-{\ml \over 2})
\]] } \ .
\eqno(B.17)
$$
The small $m$ limit of the second contribution to this second
moment can be worked out with the same technique. One gets:
$$
g_{12}={\de \over m} {-l\over \pi} \int_0^\infty dh
e^{-h^2} \ \
$$ $$
{ 1 \over
 e^{l^2/4} \[[
e^{-h \ml/2} \cM_0(h-{\ml \over 2}) + e^{h \ml/2} \cM_0(-h-{\ml \over 2})
\]]^2 } \ .
\eqno(B.18)
$$
We remember that from (B.5) the velocity two point correlations
between points at distance $l \Delta \sqrt{2} $
equals $(2 \nu / \delta)^2(g_{11}+g_{22})$ . It is therefore finite when
$m\to 0$. It is interesting to work out its small $l$ limit, which
gives the correlation between two points whose distance is small
with respect to the injection scale $\Delta$ but large with respect to
the dissipation scale (since we have taken the limit $Re \to \infty$ -or
$m \to 0$-
first):
$$
<v_1(x)v_1(y)> \simeq -{8\nu^2 \over \sqrt{\pi} m^2\Delta^2} \ {x-y \over
\Delta}
\eqno(B.19)
$$

We now turn to the computation of higher order moments. We are
interested in arbitrary moments of the velocity difference
$\vv(\vx)-\vv(\vy)$ projected onto the direction of $\vx-\vy$ (chosen as the
`$1$'
direction. The starting point is analogous to (B.4):
$$
\overline{ v_1(\vx)^p v_1(\vy)^{p'} } =
 ( {2 \nu \over \delta } )^{p+p'} \bE_{r,f}
\sum_{\a_1,...,\a_{p}}\sum_{\b_1,...,\b_{p}} e^{-(f_{\a_1}+...+f_{\a_{p}}
+f_{\b_1}+...+f_{\b_{p'}})}
$$ $$
(\vx-{\vec r}_{\a_1})_1...(\vx-{\vec r}_{\a_p})_1 \exp \((
-{(\vx-\vec r_{\a_1})^2+...+(\vx-\vec r_{\a_p})^2 \over 2} \))
$$ $$
(\vy-\vec r_{\b_1})_1...(\vy-\vec r_{\b_{p'}})_1
\exp \((
-{(\vy-\vec r_{\b_1})^2+...+(\vy-\vec r_{\b_{p'}})^2
\over 2} \))
$$ $$
\[[ \sum_\a e^{-f_\a-(\vx-\ra)^2/2\de} \]]^{-p}
\[[ \sum_\b e^{-f_\b-(\vy-\rb)^2/2\de} \]]^{-p'}
\eqno(B.20)
$$
We shall proceed as for the second moment. The first step is to
exponentiate the denominator in (B.20) using
$$
{1 \over A^p B^{p'}} = \int_0^\infty d\l \int_0^\infty d\m
{\l^{p-1} \over \Gamma(p)}
 {\m^{p'-1} \over \Gamma(p')} e^{-(\l A+\m B)} \ .
\eqno(B.21)
$$
In order to perform the averages over the values of the
free energies $f_\a$ and the positions $r_\a$ in (B.20)
we need to distinguish how many indices in $
\a_1,...,\a_{p},\b_1,...,\b_{p'}$ are distinct one from another. Let
us suppose that there appear in this sequence $k$ different
indices, which we call $\g_1,...,\g_k$. We shall call $q_j$ the
number of indices in $\a_1,...,\a_{p}$ which are equal to $\g_j$,
and $q_j'$ the
number of indices in $\b_1,...,\b_{p'}$ which are equal to $\g_j$.
Such a configuration of indices, characterized by the
number $k \in \{1,2,...,p+p'\}$, and the sequences $q_1,...,q_k,q_1',
...q_k'$ (such that $q_1+...+q_k=p$, $q_1'+...+q_k'=p'$ and for
every $j$: $q_j+q_j' \ge 1$) appears a certain number of times,
which we call
$\cC_{q_1,...,q_k,q_1',...q_k'}^{(k)}$, in the sum over $
\a_1,...,\a_{p},\b_1,...,\b_{p}$ of (B.20). For each configuration
 of indices, we just do the same transformations as for the second
moment. We shall not repeat them here, and just give the final result
which generalizes (B.12):
$$
\overline{ v_1(x)^p v_1(y)^{p'} } =
( {2 \nu \over \delta } )^{p+p'}
( {\de \over m } )^{(p+p')/2} {1 \over m}
\int_{-\infty}^\infty dh
{e^{-p'h \ml /m} \over \Gamma(p) \Gamma(p')}
\sum_{k=1}^{p+p'} \Gamma(k) m^{k-1}
$$ $$
\sum_{q_1,...,q_k,q_1',...q_k'}
 \cC_{q_1,...,q_k,q_1',...q_k'}^{(k)}
\prod_{j=1}^k \[[ {\Gamma(q_j+q_j'-m)  I_{q_j,q_j'}(h)
\over I_{0,0}(h)  } \]] \ .
\label{gf1}
\eqno(B.22)
$$
Let us now work out the leading behaviour of this expression
in the small $m$ (large $Re$) limit. From the integrals
$I_{q_j,q_j'}(h)$ we get a factor $\exp(h \ml \sum_j q_j'/m)$
which exactly compensates the explicit $e^{-p'h \ml /m}$. Using
$\de=2 m \Delta^2$, we find that the contribution to (B.22)
from a given value of $k$ scales as $ (\nu/\de)^{p+p'} (\de/m)^{-
(p+p')/2} m^{a(k)}$. A close look at the behaviours
of the integrals $I_{q,q'}$ shows that $a(k)$ is zero for $k=1,2$,
and it is strictly positive for $k \ge 3$. So we can neglect
the terms with $k \ge 3$ in (\ref{gf1}).
The term $k=1$ is easily worked out. The only allowed
configuration of indices has $q_1=p$ and $q_1'=p'$, and
its degeneracy is $\cC_{p,p'}^{(1)}=1$. For the $k=2 $
term we keep only the leading terms which correspond
to $q_1=0, \ q_2=p, \ q_1'=p', \ q_2'=0$ or to
$q_1=p, \ q_2=0, \ q_1'=0, \ q_2'=p'$, which also have a degeneracy 1
(*** a verifier ***). Altogether the leading contribution to
this moment at large $Re$ is:
$$
\overline{ v_1(x)^p v_1(y)^{p'} } =
( {2 \nu \over \delta } )^{p+p'}
( {\de \over m } )^{(p+p')/2} \mbox{sign}(l)^{p+p'}  \int dh\ \ \
$$ $$
\[[
\(( {\ml \over 2}+h\))^p \(( -{\ml \over 2}+h\))^{p'}
 {e^{-(h^2/2+l^2/8)} \over \sqrt{2\pi}} R(h,\ml) \right.
$$ $$
\left.
+(-1)^p \ml
\cM_p(-h-{\ml \over 2}) \cM_{p'}(h-{\ml \over 2}) R(h,\ml)^2
 \]] \ .
\eqno(B.23)
$$

It is a simple exercise to sum these moments and get the q$^{th}$
moment of the longitudinal velocity difference:
$$
\overline{ \((v_1(x)- v_1(y)\))^{q} } =
( {2 \nu \over \delta } )^{q}
( {\de \over m } )^{q/2} \mbox{sign}(l)^{q}   \int dh\ \
\[[
 \ml^q
 {e^{-(h^2/2+l^2/8)} \over \sqrt{2\pi}} R(h,\ml) \right.
$$ $$
\left.
+(-1)^q \ml
\int_{-h-\ml/2}^{\infty} {dt \over \sqrt{2\pi}} e^{-t^2/2}
\int_{h-\ml/2}^{\infty} {ds \over \sqrt{2\pi}} e^{-s^2/2}
(t+s)^q R(h,\ml)^2 \]] \ ,
\eqno(B.24)
$$
where:
$$
R(h,l) \equiv \left[e^{-h l /2} {\cal M}_0(h-l/2) +  e^{h l/2}
{\cal M}_0(-h-l/2)\right]^{-1}
\eqno(B.25)
$$
Under this form the moments can be inverted and we obtain the
explicit form of the pdf $P(u)$ of the velocity
difference $u \equiv v_1(x)- v_1(y)$ between two points
at distance $x-y=l \Delta \sqrt{2}$ at large
$Re$:
$$
P(u)=\de\((u-l  {2 \nu \over \delta }
\sqrt{ {\de \over m } }\)) e^{-l^2/8}
\int_{-\infty}^{\infty} {dh \over \sqrt{2\pi}}
e^{-h^2/2}   R(h,\ml)
$$ $$
+\ml \int_{-\infty}^{\infty} dh
\int_{-h-\ml/2}^{\infty} {dt \over \sqrt{2\pi}} e^{-t^2/2}
\int_{h-\ml/2}^{\infty} {ds \over \sqrt{2\pi}} e^{-s^2/2} \ \ \ \
$$ $$
 \ \ \
\de \(( u +(t+s){2 \nu \over \delta } \sqrt{ {\de \over m } } \mbox{sign}(l)
\))
R(h,\ml)^2  \ .
\eqno(B.26)
$$

\vfill
\eject

{\bf Appendix C: Correlation identities}

\vskip 1cm

In this Appendix we check that the solution we have found
does satisfy some necessary identities of correlation
functions, in the limit of large Reynolds numbers. We shall perform the
explicit check in the case of the "energy" balance. From Burgers'
 equation one finds:
$$
{1 \over 2}  { \partial \over \partial t}\overline{ ( v^\mu v^\mu)}
= A+B+C \ ,
 $$ $$
A \equiv -{1 \over 2}  \ \overline{v^\mu \partial^\mu (v^\rho v^\rho)}
\ \
B \equiv \nu \ \overline{v^\mu \partial^\mu \partial^\rho v^\rho}
\ \
C \equiv \overline{f^\rho v^\rho} \ ,
\eqno(C.1)
$$
with a convention of summation on the repeated vector indices $\mu,
\rho$ from 1 to $N$ which will be used in this whole Appendix.
The overbar denotes the expectation value
with respect to various realizations of the force $\vf$.
 We shall compute successively the three
contributions $A,B$ (from our solution), and $C$
(from a direct computation of the energy injected).
In the end we shall check that their sum
vanishes at large Reynolds number, as implied by the
stationnarity of the forced flow.

  We use the Ansatz
for the velocity given in (4.1) and developped in Appendix B.
 It will be useful to introduce
somewhat more compact notations and define:
$$
u^\mu \equiv x^\mu-r^\mu
\eqno(C.2)
$$
where $ r^\mu$ is a random variable which takes value
$r_\alpha^\mu$ with a probability $W_\alpha \exp
(-(\vx-\vra)/2\delta)$. We shall denote as before by
 brackets the expectation values with respect to this
process with fixed $W_\alpha$ and $ \vra$, while
the  averaging over $W_\alpha$ and
$ \vra$  (corresponding to the overline in (C.1))
will be denoted as in Appendix B by $E_{r,f}$.
(In terms of the directed polymer, $<O>$ denotes a thermal
average, and the overline denotes an average over quenched disorder).
With these definitions we  have:

$$
v^\mu={2\nu \over \delta} <u^\mu>
 $$ $$
\partial^\rho v^\mu={2\nu \over \delta} \((
\delta^{\rho \mu}-{1 \over \delta}(<u^\rho u^\mu>
-<u^\rho> < u^\mu>) \)) \ .
\eqno(C.3)
$$
which leads to the following  expression of $A$:

$$
A=\(( {2\nu \over \delta} \))^3  E_{r,f} \[[
-{<u^\mu>< u^\mu>}
+ {1 \over \delta} {<u^\rho>< u^\mu><u^\rho u^\mu>} \right.
$$ $$ \left.
- {1 \over \delta} {<u^\rho>< u^\mu><u^\rho>< u^\mu>} \]]
\ .
\eqno(C.4)
$$
The same steps give the expression for $B$:

$$
B=\(( {2\nu \over \delta} \))^3  {1 \over \delta} E_{r,f} \[[
-  {<u^\rho>< u^\mu >\((<u^\rho u^\mu>-<u^\rho>< u^\mu>\))}
\right.
$$ $$
\left.
+  {1 \over 2} {<u^\rho >< u^\mu u^\mu u^\rho> }
-  {1 \over 2} {<u^\rho >< u^\mu u^\mu >< u^\rho>} \]]
\ .
\eqno(C.5)
$$
The partial cancellation leaves:

$$
A+B= \(( {2\nu \over \delta} \))^3 E_{r,f}\[[
- {<u^\mu>< u^\mu>}
+  {1 \over 2 \delta} {<u^\rho >< u^\mu u^\mu u^\rho> }
\right.
$$ $$
\left.
\ \ \
-  {1 \over 2 \delta} {<u^\rho >< u^\mu u^\mu >< u^\rho>} \]]
\eqno(C.6)
$$

We now evaluate each of the terms in this expression .
The first term is given by:
$$
E_{r,f}(<u^\mu>< u^\mu>)=
E_{r,f} \[[{
\sum_{\a ,\b} W_\a W_\b (\vx-\ra)^\mu (\vx-\rb)^\mu
e^{-{(\vx-\ra)^2+(\vx-\rb)^2 \ov 2 \de}}
\ov
\((\sum_\gamma W_\gamma e^{-(\vx-\rg)^2/2 \de}\))^2}
\]]
\eqno(C.7)
$$
Using the technique of Appendix B, one finds after some work that the
only non vanishing contribution comes from the $\a=\b$
term in  (C.7) and gives:

$$
E_{r,f}(<u^\mu>< u^\mu>)={1-m \ov m}\  \delta  \ N \ .
\eqno(C.8)
$$
The same technique can be applied
 to each term in (C.6). One obtains:
$$
E_{r,f}(<u^\rho >< u^\mu u^\mu u^\rho> )={1-m \ov m^2}
\ \de^2  \ (N^2+ 2 N)
\eqno(C.9)
$$
and:
$$
E_{r,f}(<u^\rho >< u^\mu u^\mu >< u^\rho> )={1-m \ov m^2}
\ \de^2 \ (N^2+ (2-m) N)
\eqno(C.10)
$$
The sum gives in the end:
$$
A+B=- {1 \ov 2} \(( {2\nu \over \delta} \))^3 \ {1-m \ov m}\  \delta \  N
\eqno(C.11)
$$
One should notice that the terms of order $N^2$ vanish
automatically due to the structure of the velocity field (4.1),
and independently from the value of $m$.

We now proceed to the evaluation of the last term, $C$, in
the correlation identity (C.1).
This point needs a little care because one must be more precise
about the correct prescription of the forcing term, whether it is
of Ito or Stratanovitch type. To settle this issue in a
pedestrian but safe way, we have discretized the time
in units of $\tau_0 \ll 1$. The Hopf Cole mapping can be carried
out in this case, and we deduce from this computation:
$$
C={1 \over 2} \overline{f^\mu f^\mu}= {1 \over 2} N \epsilon
\eqno(C.12)
$$

Using the value (3.12) of $\de$: $\de=2 m \Delta^2$, the relation (3.7)
 between $m$ and the Reynolds
number: $
(1-{2m \over N})^{1-{N/2}} = {1\over 2} m^3 Re^3
$, together with the definition (2.4) of $Re$: $ Re^3=\epsilon \Delta^4/\nu^3$,
we finally get:
$$
\lim_{Re \to \infty} A+B+C =0 \qquad  (\epsilon  \ \ {\mbox{fixed}})
\eqno(C.13)
$$
which establishes the correlation identity for the energy balance (C.1).

Using the
same techniques, we have also checked that the correlation identities
corresponding to
${\partial \over \partial t} \overline {\vec \nabla .\vec v} =0$
and
${\partial \over \partial t} \overline {\((( \vec v)^2\))^2} =0$
also hold {\it for any N} in the limit of large $Re$.

\vfill
\eject

{\bf Appendix D: Replica computation of the velocity two points correlation}

\vskip 1cm

The aim of this Appendix is to
 compute the two point correlation $\overline{v_(x,t)v(y,t)} $
directly from the replica method of sect.3. This will provide a check
that the intermediate physical representation (B.1) is safe. In the course
of this computation we shall also establish useful replica identities
which can be of wider interest. To keep the computations simple we consider
only the one dimensional case, $N=1$. We start from the result (3.8)
for the partition function for $n$ replicas of the polymer arriving
at time $t$ at the points $\vx_a,a=1,...,n$:
$$
\overline{Z(x_1,t)...Z(x_n,t)} = c \sum_\pi
 \exp \(({1\over 2} \sum_{a,b} R_{\pi(a)\pi(b)} x_a(t)
x_b(t) \))
\eqno(D.1)
$$
We shall compute $\overline{Z(x,t)^{n/2} Z(y,t)^{n/2}}$ which can be written
as:
$$
\overline{Z(x,t)^{n/2} Z(y,t)^{n/2}}= c \sum_{ \{ \sigma \} }' \ \ \ \
$$
$$
 \exp \(({1\over 2} \sum_{a,b} R_{ab} \left[ {1-\sigma_a \over 2} x +
{1 + \sigma_a \over 2} y \right]
 \left[ {1-\sigma_b \over 2} x +
{1 + \sigma_b \over 2} y \right]
  \))
\eqno(D.2)
$$
where the $\sum_{ \{ \sigma \} }'$ is over $n$ Ising spins $\sigma_a=\pm 1$,
with
the constraint that $\sum_a \sigma_a =0$.

As an intermediate step to this computation, we first compute the
action of $\sum_{ \{ \sigma \} }'$ onto a polynomial in the spins. Let
$a_1,...,a_k$ be $k$ different replica indices. We define:
$$
A_k \equiv \sum_{ \{ \sigma \} }' \sigma_{a_1}.. \sigma_{a_k} \ .
\eqno(D.3)
$$
Clearly $A_1=0$. As for $A_2$, one can use $0=\sum_{a,b} \sum_{ \{ \sigma \} }'
\sigma_a
\sigma_b = n+ n(n-1) A_2$ to deduce $A_2=1/(1-n)$. The general result can
be deduced from an iteration of the above procedure:
$$
A_k =0  \ \ \ \ \ \ \ \ \ \ \ \hbox{if k is odd}
$$
$$
A_k= {1 \over 1-n} {3 \over 3-n} ... {k-1 \over k-1-n} \ \ \ \hbox{if k is
even}
\eqno(D.4)
$$
and this can be written in the form:
$$
A_k = {2^{n+1} \over B(-n/2,-n/2)} \int_{-\infty}^{\infty} dy \ (cosh(y))^n/2
(tanh(y))^k
\eqno(D.5)
$$
This expression is well defined for $n<0$, and can be continued analytically
to positive $n$. $B$ is Euler  beta function \cite{GraRy}. From this expression
of $A_k$ one can deduce the generating function:
$$
g[h_1...h_n] \equiv \sum_{ \{ \sigma \} }' \exp \(( \sum_a h_a \sigma_a \))
= (\prod_a cosh(h_a)) .
$$
$$
 \sum_{k=0}^n A_k \sum_{a_1<...<a_k} tanh(h_{a_1})...tanh(h_{a_k})
$$
$$
\ \ \ \ = {2^{n+1} \over B(-n/2,-n/2)} \int_{-\infty}^{\infty} dy \
\prod_{a=1}^n cosh(y+h_a)
\eqno(D.6)
$$
Formula (D.6) is a replica identity which may turn out useful in
other contexts. Here we will use it to compute $\overline{v(x)v(y)}$.
Starting from (D.2), and using the fact that $\sum_b R_{ab}=0$ (derived in
Appendix A), we have:
$$
\overline{Z(x,t)^{n/2} Z(y,t)^{n/2}}= c \sum_{ \{ \sigma \} }'
\exp\(({\a \over 2} \sum_{a,b} R_{ab}\sigma_a \sigma_b \))
\eqno(D.7)
$$
where $\a$ is defined as:
$$
\a={(x-y)^2 \over 4}
\eqno(D.8)
$$
We call as usual $\tilde r, r_1, r_0$ the various elements of the hierarchical
$R_{ab}$ matrix. Using the fact that $r_0=0$, we get from (D.6):
$$
\overline{Z(x,t)^{n/2} Z(y,t)^{n/2}}= {2^{n+1} \over B(-n/2,-n/2)}
\exp \(( \a {n \over 2} (\tilde r -r_1) \))
$$
$$
 \int_{-\infty}^{\infty} dh_0
\[[\int Dh \cosh^m(h_0+h\sqrt{\a r_1}) \]]^{n/m}
$$
$$
= {2^{n+1} \over B(-n/2,-n/2)}
\exp \(( \a {n \over 2} (\tilde r -r_1+m r_1) \))
$$
$$
 \int_{-\infty}^{\infty} dh_0 e^{n h_0}
\[[\int Dh \((1+e^{-2 h_0-2 h\sqrt{\a r_1})-2 m \a r_1}\))^m \]]^{n/m}
\eqno(D.9)
$$
where $Dh \equiv dh/\sqrt{2 \pi} \exp(-h^2/2)$. Changing variables to
$\mu=\exp(-2 h_0)$ and $h= m \sqrt{r_1} (z-x) \mbox{sign}(x-y)$, and using the
fact that
$\tilde r -r_1+m r_1=0$, we derive:
$$
\overline{Z(x,t)^{n/2} Z(y,t)^{n/2}}=
 {2^{n} \over B(-n/2,-n/2)}
\int_0^\infty {d\mu \over \mu} \mu^{-n/2} \ \ \
$$
$$
\ \ \ \
\[[{m\sqrt{r_1} \over 2 \pi} \int dz \((e^{-m r_1 (z-x)^2/2}
+\mu e^{-m r_1 (z-y)^2/2}\))^m \]]^{n/m} \ .
\eqno(D.10)
$$
Having computed $\overline{Z(x,t)^{n/2} Z(y,t)^{n/2}}$, the velocity
correlation is easily deduced as:
$$
\overline{v(x)v(y)}= 4 \nu^2 \overline{
\(( {Z'(x) \over Z(x)} {Z'(y) \over Z(y)} \)) }
=\lim_{n \to 0} {16 \nu^2 \over n^2}
{\partial^2 \over \partial x \partial y}
\overline{Z(x,t)^{n/2} Z(y,t)^{n/2}}
\eqno(D.11)
$$
It is straightfoward to check that this  replica computation agrees with the
direct physical space computation of Appendix B. More precisely we
recover the expression (B.5),(B.10),(B.11) for the correlation, using the
fact that $m r_1 \delta =1$.

\vskip 1cm

\centerline{\bf{Figure Captions}}

\vskip 1cm

Fig. 1: Typical snapshots of the velocity field as given by Eq. (4.1), with $m
= {1 \over Re} = 10^{-2}$, in one dimension $N=1$ (Fig. 1-a), or two dimensions
$N=2$ (Fig 1-b, where we have plotted in grey levels the modulus of the
velocity field).

Fig 2: Comparison of the numerically determined velocity correlation function
(directly from Eq. (4.1)), and our analytical formula, obtained after
integrating Eq. (4.7), multiplied by $u^2$. Note the linear regime at small
$x-y$.

Fig 3. Sketch of the `multifractal' spectrum $\zeta(q)$, giving the $r$
dependence of the $q$ th moment of the velocity field, both for the forced
Burgers' equation (triangles) and hydrodynamical turbulence (circles - see,
e.g. \cite{She}). Note that $\zeta(3)=1$ for both models (Kolmogorov's
scaling). Intermittency corrections (i.e. the departure of $\zeta(q)$ from $q
\over 3$, as given by the dashed line) are much stronger in Burgers'
turbulence: this is due to the fact that singularities are concentrated on
$N-1$ dimensional structures, rather than on vorticity tubes.

\bibliographystyle{unsrt}

\vfill
\eject
\begin{figure}
\begin{center}
\leavevmode
\epsfysize=350pt\rotate[r]{\hbox{\epsfbox[107 3 586 757]{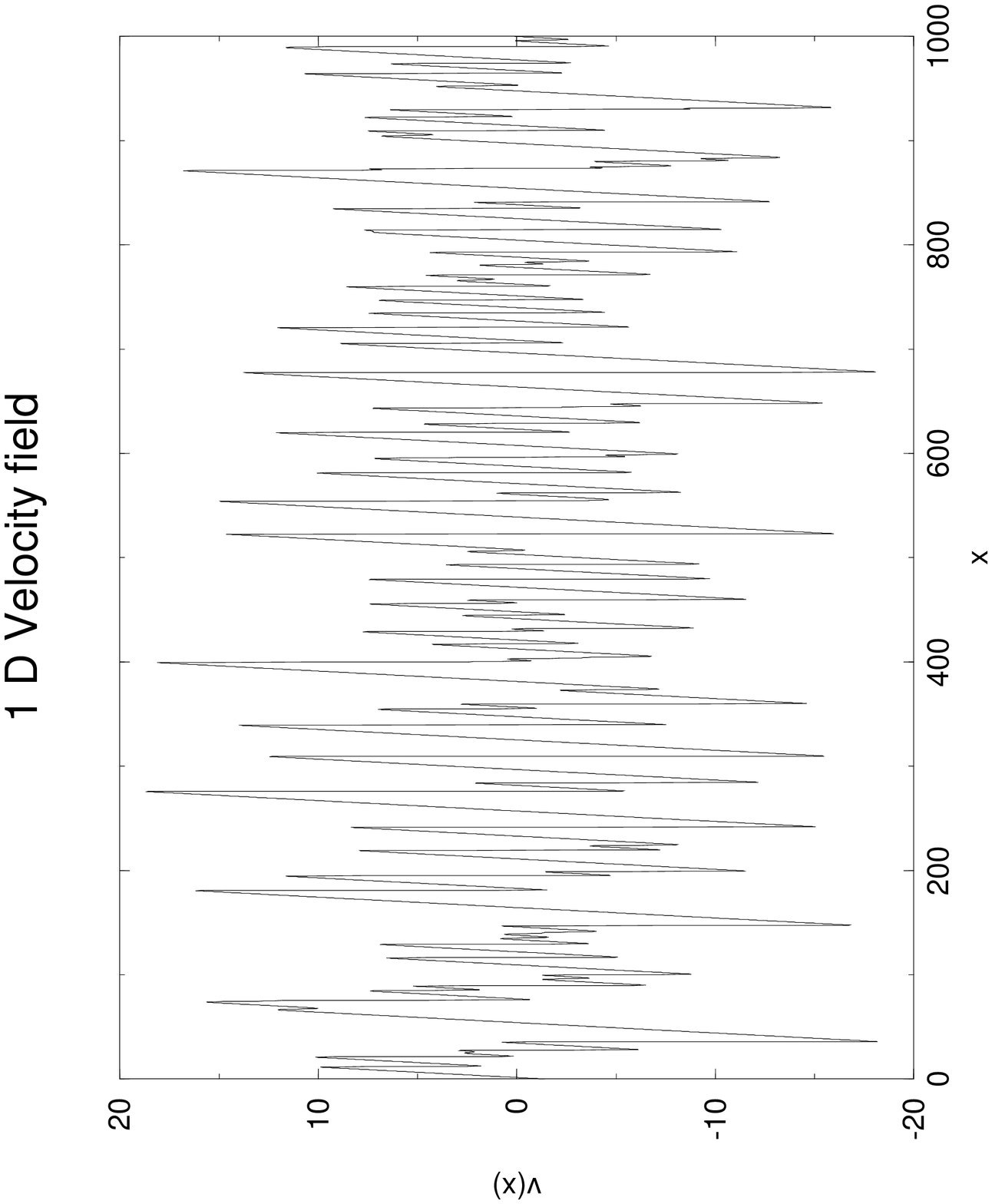}}}
\end{center}
\end{figure}

\vfill
\eject
\begin{figure}
\begin{center}
\leavevmode
\epsfbox{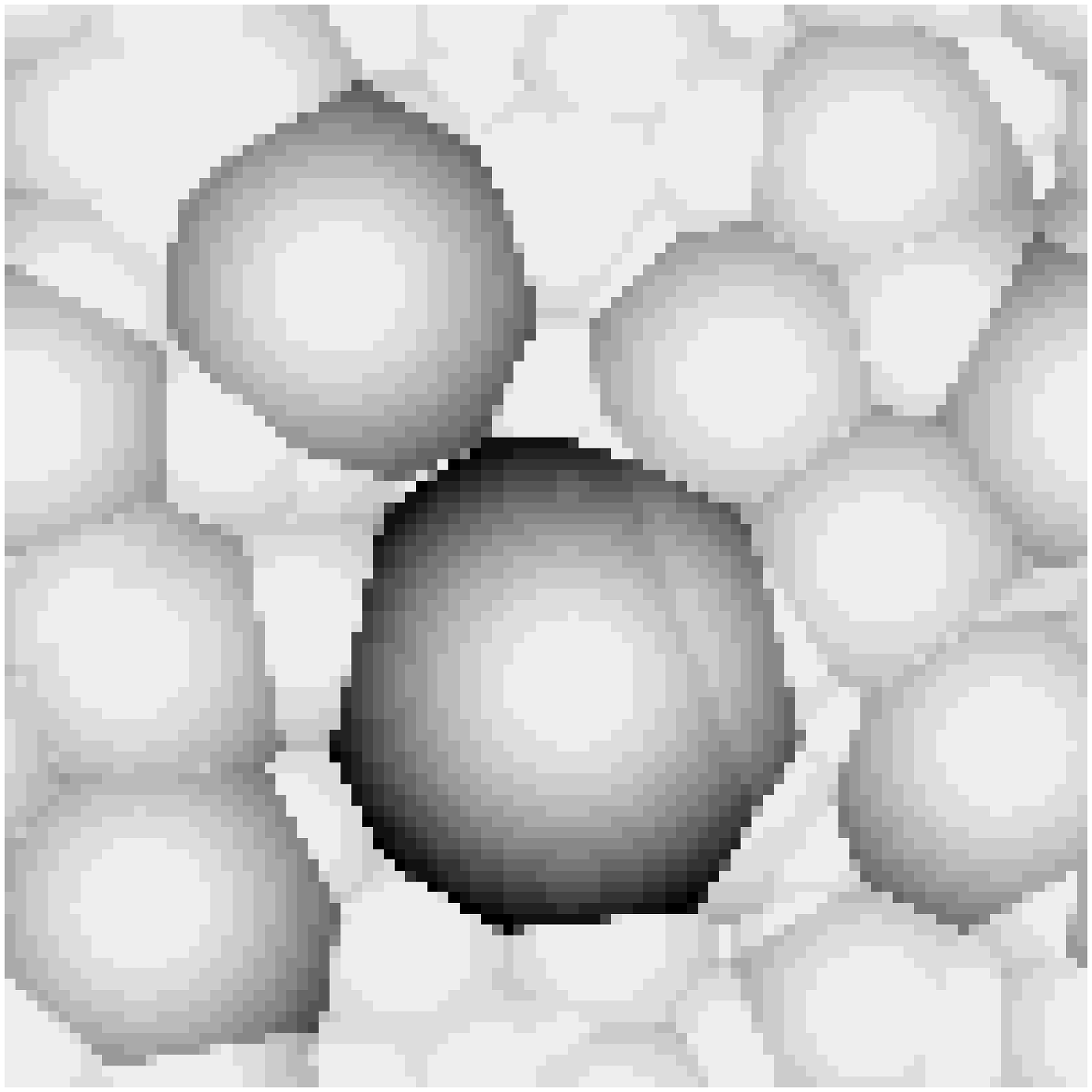}
\end{center}
\end{figure}

\vfill
\eject
\begin{figure}
\begin{center}
\leavevmode
\epsfysize=350pt\rotate[r]{\hbox{\epsfbox[107 3 586 757]{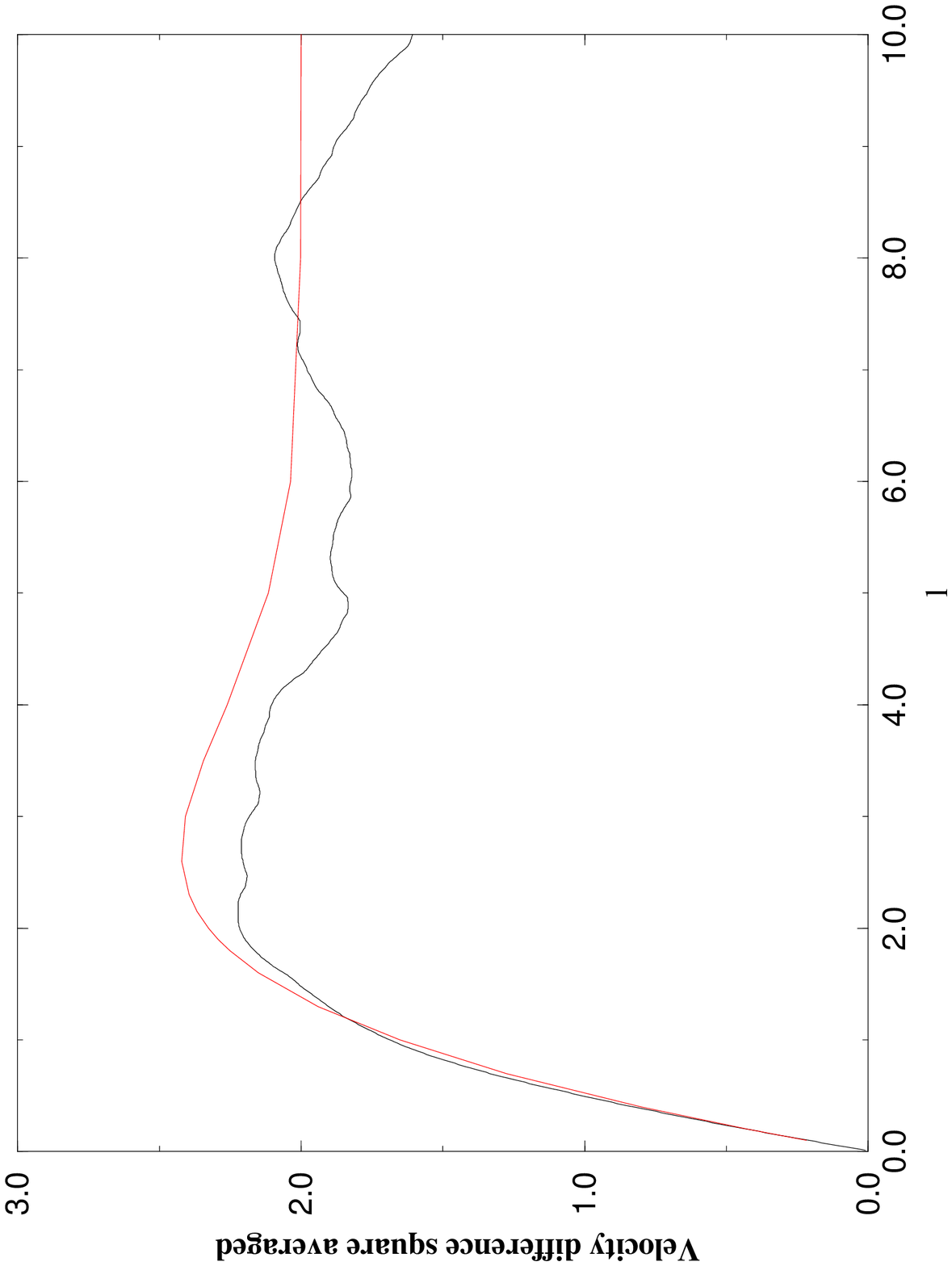}}}
\end{center}
\end{figure}

\vfill
\eject
\begin{figure}
\begin{center}
\leavevmode
\epsfysize=350pt\rotate[r]{\hbox{\epsfbox[107 3 586 757]{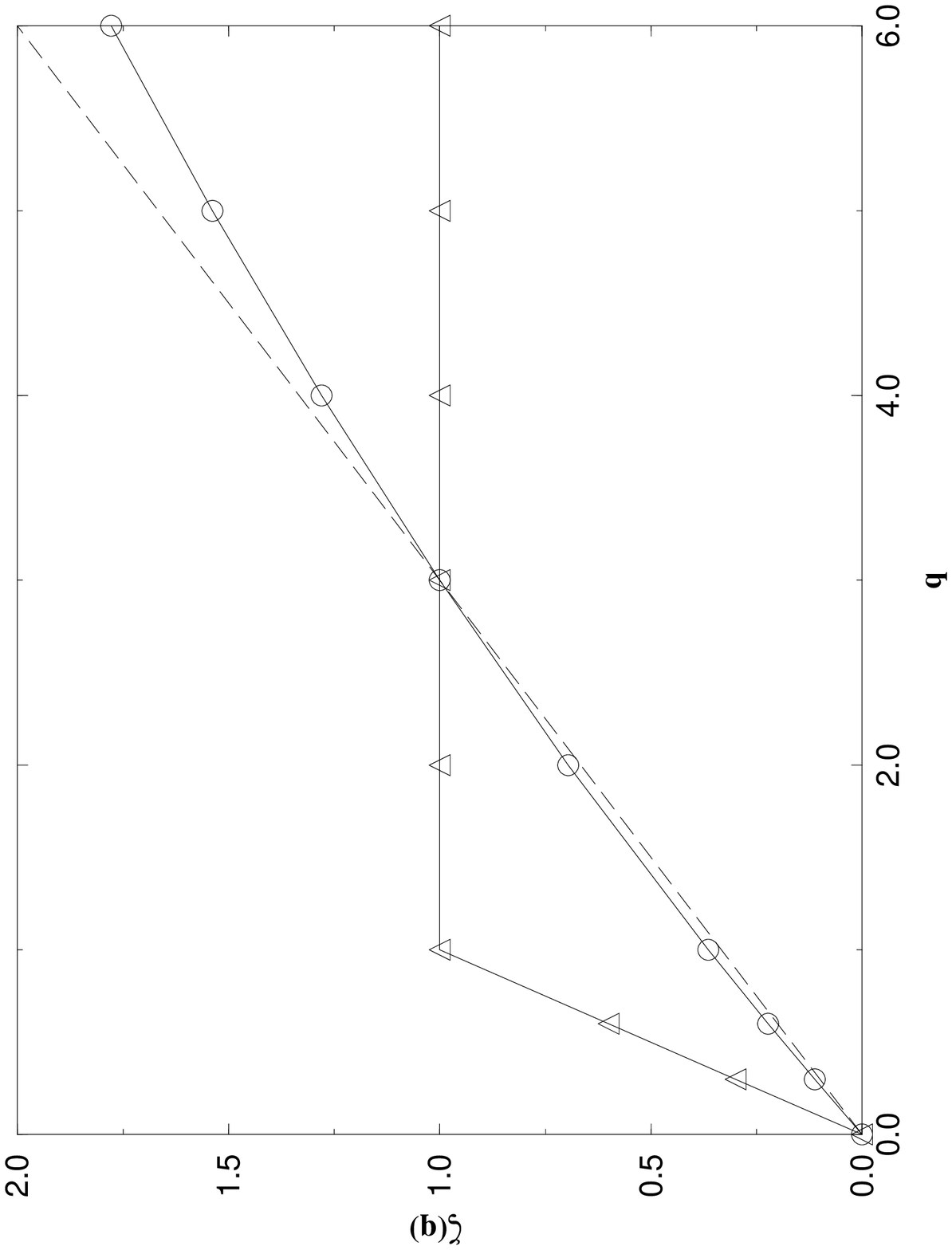}}}
\end{center}
\end{figure}

\end{document}